\documentclass[aps,reprint,a4paper]{revtex4-1}

\usepackage{graphicx}
\usepackage{dcolumn}
\usepackage{bm}
\usepackage{amsmath}
\usepackage{array}
\usepackage{color}
\usepackage{ulem}
\usepackage{longtable}
\usepackage{float}
\usepackage{afterpage}

\setcitestyle{numbers,square}

\usepackage{soul}

\begin{document}

\preprint{APS/123-QED}

\title{Locating the eigenshield of a network via perturbation theory}

\author{Ming-Yang Zhou$^1$}

\author{Manuel Sebastian Mariani$^{2,3}$}

\author{Hao Liao$^1$}%

\author{Rui Mao$^1$}

\author{Yi-Cheng Zhang$^4$}
\affiliation{$^1$College of Computer Science and Software Engineering, Shenzhen University, Shenzhen, 518060, P. R.  China}
\affiliation{$^2$Institute of Fundamental and Frontier Sciences, University of Electronic Science and Technology of China, Chengdu 610054, P.R.  China}
\affiliation{$^3$URPP Social Networks, University of Zurich, CH-8050 Zurich, Switzerland}
\affiliation{$^4$Physics Department, University of Fribourg, Chemin du Mus$\acute{e}$e 3, 1700 Fribourg Switzerland}

%
%

\date{\today}

\begin{abstract}
The functions of complex networks are usually determined by a small set of vital nodes.
Finding the best set of vital nodes (eigenshield nodes) is critical to the network's robustness against rumor spreading and cascading failures, which makes it one of the fundamental problems in network science.
The problem is challenging as it requires to maximize the influence of nodes in the set while simultaneously minimizing the redundancies between the set's nodes. However, the redundancy mechanism is rarely investigated by previous studies.
Here we introduce the matrix perturbation framework to find a small ``eigenshield" set of nodes that, when removed, lead to the largest drop in the network's spectral radius.
We show that finding the ``eigenshield" nodes can be translated into the optimization of an objective function that simultaneously accounts for the individual influence of each node and redundancy between different nodes.
 We analytically quantify the influence redundancy that explains why an important node might play an insignificant role in the ``eigenshield" node set.
Extensive experiments under diverse influence maximization problems, ranging from network dismantling to spreading maximization, demonstrate that the eigenshield detection tends to significantly outperforms state-of-the-art methods across most problems. Our findings shed light on the mechanisms that may lie at the core of the function of vital nodes in complex network.

\end{abstract}

\maketitle

\section{Introduction}

A core problem in the physics of complex systems concerns the identification of ``vital" nodes that play a fundamental role in the structure and dynamics of complex networks, e.g., for population immunization, spreading maximization in epidemic processes, optimal network percolation~\cite{morone2015influence,lu2016vital,pei2020influencer}. Finding the vital nodes (eigenshield nodes) for a network's structural robustness can help optimize vaccination strategies~\cite{wang2016statistical}, prevent the collapse of infrastructure systems~\cite{wang2009cascade} and ecosystems~\cite{saavedra2011strong,dominguez2015ranking}.
In parallel, from a dynamic perspective, determining the eigenshield nodes has far-reaching implications for viral marketing campaigns~\cite{kempe2003maximizing} and epidemic spreading processes~\cite{kitsak2010identification,wong2020evidence}.
The search for eigenshield nodes can be translated into well-defined ``influence maximization problems" (IMPs)~\cite{lu2016vital,de2014role,radicchi2017fundamental,iannelli2018influencers,aral2018social,mariani2020network}.
Solving IMPs typically requires not only to maximize the influence of the nodes included in the eigenshield node set, but also to simultaneously minimize their redundancy~\cite{lu2016vital,ji2017effective}.
Yet we still lack an analytic method to quantify the redundancy of a given node set and its relationship with the detection of optimal eigenshield nodes.


Here we introduce a theoretical framework to find a small ``eigenshield" set of nodes that, when removed, lead to the largest drop in the network's spectral radius.
The rationale behind this problem is that the spectral radius plays a key role for diverse structural and dynamical properties of complex networks, including the epidemic threshold~\cite{chakrabarti2008epidemic}, linear threshold dynamics~\cite{kempe2003maximizing}, and network robustness~\cite{staniczenko2013ghost}. As a direct consequence, we expect the eigenshield nodes to play a fundamental role for diverse structural properties and dynamical processes on the network.

We use network perturbation theory~\cite{restrepo2006characterizing} to map the eigenshield detection problem into
the optimization of an objective \textit{eigenshield} function, which features two components: (a) a positive ``influence" contribution, which represents the sum of the influences the eigenshield nodes would have if they were to be considered independently, and (b) a negative ``redundancy" contribution, which represents the redundancy term (``overlapping influence", \textit{OI} for short) that results from the nodes' underlying interactions inside the eigenshield node set.
We show that the previously-neglected redundancy term explains why a central node according to traditional centrality metrics~\cite{albert2004structural,krzakala2013spectral,mugisha2016identifying,lu2016vital} might play an insignificant role in the eigenshield node set. We then develop an optimization method to optimize the eigenshield function, and to identify the eigenshield nodes.

Analysis of $40$ empirical networks
reveals that our method outperforms state-of-the-art (heuristic and deep learning based)
centralities~\cite{morone2015influence,lu2016vital,albert2004structural,page1999pagerank,restrepo2006characterizing,kitsak2010identification,mugisha2016identifying,krzakala2013spectral}
 in both the spectral radius minimization problem and a set of other IMPs-- including the problem of the dismantling of a network's giant component~\cite{morone2015influence,ren2019generalized}, maximizing the spreading coverage in the linear threshold model~\cite{kempe2003maximizing}, the susceptible-infected-susceptible (SIS) model, and the susceptible-infected-recovered (SIR) model~\cite{kitsak2010identification}). 
 Unlike the eigenshield nodes identified by the state-of-the-art methods, the eigenshield nodes identified by the proposed method exhibit large cumulative influence and small overlapping influence. Eigenshield nodes detected by the state-of-the-art methods can exhibit a large cumulative influence; however, their redundancy is also high, which can degrade their performance.
Taken together, these findings point to the large redundancy between the detected nodes as the main drawback of the state-of-the-art methods, and indicate the removal of redundant nodes from the eigenshield set as a viable pathway to overcoming this limitation.

\section{Results}

%


\subsection{Set influence }
We consider an arbitrary structural perturbation of a symmetrical $N\times N$ adjacency matrix, $\mathsf{A}$, whose elements are denoted as $a_{ij}$, $a_{ij}=1$ if there is an edge between node $i$ and $j$, and $a_{ij}=0$ otherwise. We parametrize the perturbation as $\mathsf{A}'=\mathsf{A}+\varepsilon \mathsf{P}$, where $\varepsilon$ denotes an arbitrary small real number, and $\varepsilon \mathsf{P}$ denotes an arbitrary perturbation matrix. The terms $\{\mu_i\}$ and $\{\mathbf{v}_i\}$ denote the eigenvalues and corresponding eigenvectors of $\mathsf{A}$ (with $\mu_1\geq \mu_2\geq...\geq \mu_N$, $||\mathbf{v}_i||_2=1$).
The eigenvalues $\{\mu'_i\}$ of the perturbed matrix, $\mathsf{A}'$, can be approximated by $\mu_i'\approx\mu_i+\varepsilon (\mathbf{v}_i^TP\mathbf{v}_i)$ ~\cite{restrepo2006characterizing} (See Appendix E  for the derivation).

We interpret the removal of a set of nodes from a given network as a perturbation. Particularly, removing a set $\mathcal{S}$ of nodes causes the loss of all edges $\mathcal{E}_{\mathcal{S}}=\{(i,j)|i\in\mathcal{S} \text{ or } j\in\mathcal{S}\}$ attached to nodes in $\mathcal{S}$. Hence, the removal of the nodes in $\mathcal{S}$ can be represented by an $N\times N$ perturbation matrix $\mathsf{R}$, whose element $r_{ij}=a_{ij}$ if $i\in \mathcal{S}$ or $j\in \mathcal{S}$, and $r_{ij}=0$ otherwise.
After removing the edges attached to the node set, the adjacency matrix of the remaining network is $\mathsf{A}'=\mathsf{A}-\mathsf{R}$.
By assuming that $|\mathcal{S}|\ll N$ and replacing $\varepsilon=-1$ and $\mathsf{P}=\mathsf{R}$, we obtain $\mu_i'\approx \mu_i-\Delta\mu_i$ for $\mu'_i$ ($i=1,2,\dots,N$), with
\begin{equation}
\label{eq:decrease}
\Delta\mu_i(\mathcal{S})=\mathbf{v}_i^T \,\mathsf{R}\,\mathbf{v}_i=2\,\sum_{(a,b)\in\mathcal{E}_{\mathcal{S}}}v_{ia}\,v_{ib}=\sum_{(a,b)\in\mathcal{E}_{\mathcal{S}}}s_{iab},
\end{equation}
where we define the link-level score $s_{iab}=2\,v_{ia}\,v_{ib}$, representing the decrease in eigenvalue $u_i$ induced by removing the edge $(a,b)$, and $v_{ia}$ denotes the $a-$th entry of the eigenvector $\mathbf{v}_i$.
We refer to Eq.~\eqref{eq:decrease} as the \textit{set influence} of $\mathcal{S}$.
The term $\Delta\mu_i(\mathcal{S})$ is the sum of the eigenvalues decreased by the removal of every edge attached to the removed nodes.
 Notably, we only consider the first-order approximation of $\mu_i'$.
The first-order approximation of $\mu'_i$ is sufficiently accurate when a small fraction of nodes is removed (See Appendix Fig. \ref{fig:Approximatereallambda} for the empirical illustration).

To reveal the relation between the set influence and influence of the individual node, we note that if we remove a single node $u$, the decrease in eigenvalues is as follows:
\begin{equation}
  \Delta\mu_i(u)=\sum_{v}a_{uv}\,s_{iuv},
  \label{eq:singleinfluence}
\end{equation}
which we refer to as the \textit{node influence}.
However, the set influence $\Delta\mu_i(\mathcal{S})$ is not simply the sum of the influence scores of the nodes in $\mathcal{S}$. The set influence can also be expressed as follows:
\begin{equation}
    \Delta\mu_i(\mathcal{S})=
    \Delta\mu_{i,\text{cum}}(\mathcal{S})-\Delta\mu_{i,\text{ov}}(\mathcal{S}),
    \label{influence}
\end{equation}
where $\Delta\mu_{i,\text{cum}}(\mathcal{S}):=\sum_{u\in\mathcal{S}}\Delta\mu_i(u)$ represents the \textit{cumulative influence} of the nodes in $\mathcal{S}$, considered independently, and $\Delta\mu_{i,\text{ov}}(\mathcal{S})$ represents the \textit{OI} of the nodes in $\mathcal{S}$, defined as follows:
\begin{equation}
\label{eq:setandoverlapinfluence}
  \Delta\mu_{i,\text{ov}}(\mathcal{S})=  \sum_{(a,b)\in\mathcal{E}^{\text{int}}_{\mathcal{S}}}s_{iab},
\end{equation}
where $\mathcal{E}^{\text{int}}_{\mathcal{S}}=\{(i,j)|i\in\mathcal{S} \text{ and } j\in\mathcal{S}\}$. We refer to Section IV for the derivation of Eq.~\eqref{influence}.
Intuitively, when we calculate the set influence, each edge should be calculated only once in Eq. \eqref{eq:decrease}. However, the \textit{cumulative influence} term counts
the edges between nodes in $\mathcal{S}$ twice, with one extra contribution removed from the set influence function (the OI term in Eq.~\eqref{influence}).

Eq. \ref{eq:singleinfluence} describes the scenario that $\mathcal{S}$ only has a single node $u$. With the increase of $|\mathcal{S}|$, the  contribution of node $u$ to the set influence $\Delta\mu_i(\mathcal{S})$ is
\begin{equation}
\label{eq:eigeninfluence}
\widetilde{\Delta\mu_i}(u)=\Delta\mu_i(\mathcal{S})-\Delta\mu_i(\mathcal{S}\setminus  u)=\sum_{v\notin \mathcal{S}}a_{uv}\,s_{iuv}.
\end{equation}
The eigenshield $\widetilde{\Delta\mu_i}(u)$ is determined by both the edges attached to $u$ and the set $\mathcal{S}$. With the increase of $|\mathcal{S}|$, the neighbors of $u$ are likely to be chosen and added to $\mathcal{S}$, leading to the decrease of $\widetilde{\Delta\mu_i}(u)$. When $\widetilde{\Delta\mu_i}(u)$ is small enough, node $u$ plays an insignificant role for the set influence of $\mathcal{S}$ and should be removed from the eigenshield node set.

If the perturbation preserves the order of the eigenvalues (\textit{i.e., } if $\mu_i>\mu_j$ implies $\mu'_i>\mu'_j$), the problem of determining the set $\mathcal{S}$ that maximizes the decrease in the largest eigenvalue is equivalent to maximizing $\Delta\mu_1(\mathcal{S})$. However, in general, this was not the case. The precise formulation involves determining the set $\mathcal{S}$ that minimizes $\lambda_1(\mathcal{S})=\max_{i\in\{1,\dots,N\}}\{\mu_i-\Delta\mu_{i}(\mathcal{S})\}$. To save computational time, in practice, we can achieve satisfactory accuracy by considering only the largest $h$ original eigenvalues and minimizing
\begin{equation}
\label{eq:setinfluence}
    \tilde{\lambda}_1(\mathcal{S})=\max_{i\in\{1,\dots,h\}}\{\mu_i-\Delta\mu_{i}(\mathcal{S})\},
\end{equation}
where $h\ll N$ denotes a small positive number. There is no principled criterion for setting $h$. Increasing $h$ increases the precision at the cost of higher computational time. The empirical results suggest that a small value of $h$ is sufficient to achieve high precision. In the experiments, we set $h=20$ unless otherwise stated; the results obtained are robust with respect to variations in $h$.



\begin{figure*}[t]
  \centering
  \includegraphics[trim={0cm 0cm 0cm 0cm},clip=true, width=6in]{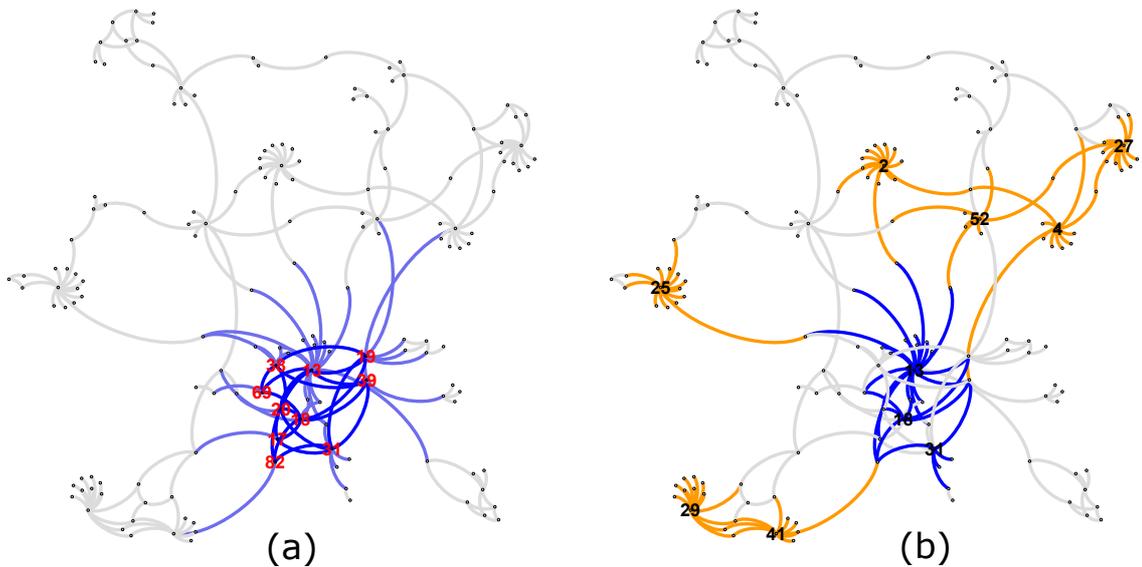}\\
  \caption{\textbf{An illustration of the detected eigenshield nodes by the eigenvector centrality (a) and the SOI method (b).}
 (a) Ten candidate eigenshield nodes (blue) detected by the eigenvector centrality (EC) in the PDZBase network. 
 The nodes are tightly clustered, which impairs their set influence.
 (b) Ten candidate nodes detected by the SOI method. Nodes detected by SOI, but not by EC are colored in yellow.
The eigenshield nodes of SOI methods are located sparsely, which implies a smaller influence overlap than that of EC.
 }
  \label{fig:PDZBaseplot}
\end{figure*}

\subsection{Eigenshield and optimization method}
We then introduce a \textit{greedy method}
to minimize $\tilde{\lambda}_1(\mathcal{S})$. We start from an empty set, $\mathcal{S}=\emptyset$, and at each step, we select and include the candidate node $u\notin \mathcal{S}$ that minimizes $\tilde{\lambda}_1(\mathcal{S}\cup \{u\})$. After the new node $u$ is added, we remove all nodes $ v \in \mathcal{S}$ from $\mathcal{S}$ that contributes to $\tilde{\lambda}_1$ less than $u$. The contribution of an individual node $u$ to $\tilde{\lambda}_1$ (eigenshield value) is defined as
\begin{equation}
\label{eq:eigeninfluence3}
\tilde{\lambda}_1(u)=\tilde{\lambda}_1(\mathcal{S}\backslash\{u\})-\tilde{\lambda}_1(\mathcal{S}).
\end{equation}
 The addition of a new node and the removal of weaker contributors are repeated until the process converges, i.e. until we obtain a fixed number of nodes.
It is noticed that $\tilde{\lambda}_1(u)$ in Eq. \ref{eq:eigeninfluence3} is degenerated into $\widetilde{\Delta\mu_1}(u)$ in Eq. \ref{eq:eigeninfluence} when we set $h=1$.

Despite its conceptual simplicity, the greedy algorithm has high time complexity because it requires the calculation of the eigenvalues and corresponding eigenvectors of the adjacency matrix $\mathsf{A}$, which has time complexity $O(N^3)$.
To overcome this issue, herein we propose a highly scalable method based on a simplified influence function. The function weighs the decreases in eigenvalues with the magnitude of every original eigenvalue, reflecting the property that the largest eigenvalue of the perturbed matrix $\mathsf{A}'$ tends to be mostly determined by the variations in the largest original eigenvalues of matrix $\mathsf{A}$.
Therefore, we introduce the weighted sum,
$
    w(\mathcal{S})=\sum_{i=1}^{N}\mu_i\,\Delta\mu_i(\mathcal{S}).
    \label{weighted}
$
Using Eq.~\eqref{eq:decrease}, $w(\mathcal{S})$ can be reduced to:
\begin{equation}
\label{eq:soiformalism}
    w(\mathcal{S})=\sum_{i\in\mathcal{S}}k_i-\frac{1}{2}\sum_{i\in\mathcal{S}} k_i^{\text{(int)}}(\mathcal{S})
\end{equation}
where $k_i^{\text{(int)}}(\mathcal{S})=\sum_{j\in\mathcal{S}}a_{ij}$ denotes the internal degree of node $i$ within the set $\mathcal{S}$. 
Based on the described analysis, the IMP problem is rephrased of determining a set of nodes $\mathcal{S}$ to maximize $w(\mathcal{S})$. We maximize $w(\mathcal{S})$ using a similar \textit{greedy algorithm} to that used to minimize $\tilde{\lambda}_1(\mathcal{S})$.
The difference is that at each step, we select a node to maximize $w(\mathcal{S})$.
In addition, the removal of previous nodes is based on the reduced eigenshield,
\begin{equation}
w(u)={w}(\mathcal{S})-{w}(\mathcal{S}\backslash\{u\})=k_i-k_i^{\text{(int)}}(\mathcal{S}).
\end{equation}
In fact, the $\tilde{\lambda}_1(\mathcal{S})$-based greedy algorithm directly optimizes the OI in Eq.~\eqref{influence}, and we refer to the corresponding set detection method as the OI method, whereas the $w(\mathcal{S})$-based greedy algorithm optimizes the simplified OI in Eq.~\eqref{eq:soiformalism}, and we refer to the corresponding set detection method as the SOI method.

Before proceeding with extensive performance validation, we focus on the major differences between the SOI and state-of-the-art methods. Similar to Eq.~\eqref{influence}, the influence function $w(\mathcal{S})$ is decomposed as the difference between a term representing the cumulative influence of the nodes in $\mathcal{S}$ ($\sum_{i\in\mathcal{S}}k_i$), and a term quantifying the degree of internal connectedness of the nodes in $\mathcal{S}$ (proportional to $\sum_{i\in\mathcal{S}} k_i^{\text{(int)}}(\mathcal{S})$). A set with high influence, $w(\mathcal{S})$, is simultaneously characterized by a large cumulative influence of its individual nodes and a low degree of internal connectedness. The latter property ensures that there is minor significant redundancy in the network paths that connect the detected nodes.
Elder significant nodes that have large redundancy with fresh nodes could be identified and removed from the eigenshield node set.

To appreciate the role of SOI on the optimal set selection, we visually compare the optimal sets by the SOI against those by the traditional EC in a specific empirical network (Fig. \ref{fig:PDZBaseplot}).
Both methods are relevant to the largest eigenvector of the adjacency matrix $\mathsf{A}$.
However, the nodes detected by the EC (Fig. \ref{fig:PDZBaseplot}a) are densely connected, whereas those detected by the SOI method (Fig. \ref{fig:PDZBaseplot}b) are sparsely connected, implying that the nodes detected by the SOI method have a smaller influence overlap than those detected by the EC.



\begin{figure}
  \centering
  \includegraphics[width=3in]{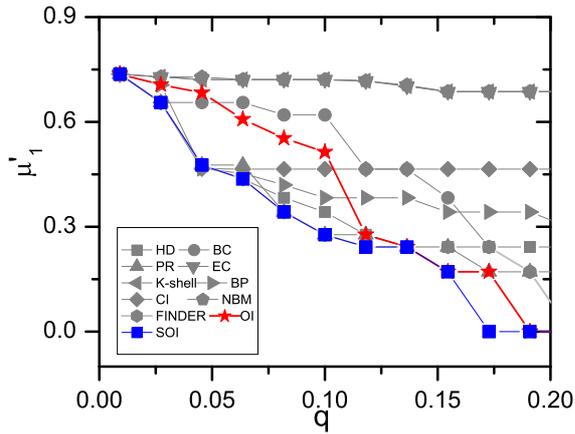}\\
  \caption{\textbf{Methods' performance in a single network.} We compare the performance of the proposed methods (OI and SOI) against those of the eight state-of-the-art methods in the PDZBase network for the \textit{eigenvalue minimization problem}.
  The figure displays the largest eigenvalue $\mu'_{1}$ as a function of the fraction $q$ of the removed nodes. Better methods exhibit a smaller area under $\mu'_{1}(q)$.
 }
  \label{fig:PDZBase}
\end{figure}

\subsection{Numeric results}
We begin by considering a single network. We compared the performance of the OI and SOI methods (to minimize $\tilde{\lambda}_1(\mathcal{S})$ and maximize $w(\mathcal{S})$, respectively) against those of nine state-of-the-art methods: high degree (HD)  \cite{lu2016vital}, betweenness centrality (BC) \cite{albert2004structural}, PageRank (PR) index~\cite{brin1998anatomy}, eigenvector centrality (EC) \cite{restrepo2006characterizing}, K-shell index~\cite{kitsak2010identification}, belief propagation index \cite{mugisha2016identifying}, collective influence (CI)~\cite{morone2015influence}, non-backtracking matrix (NBM) index \cite{krzakala2013spectral}, and FINDER index (a reinforcement learning method) \cite{fan2020finding}.
We refer to Appendix A for the details of the state-of-the-art methods.
 Figure \ref{fig:PDZBase} shows the performance of the eleven methods in the PDZBase network.
In  Fig. \ref{fig:PDZBase}, the proposed methods outperformed the other methods in terms of $\mu_{1}'$.

To further validate the methods, we consider the eigenvalue minimization problem directly addressed by the OI and SOI methods, 
as well as two well-studied IMPs: the structural problem of determining the set of nodes whose removal causes the biggest decrease in the size of the giant component (network dismantling problem~\cite{morone2015influence}), and the functional problem of determining the set of nodes that maximize the spreading of information under the linear threshold model (spreading maximization problem~\cite{kempe2003maximizing}).

Beyond analyzing a single dataset, we analyzed 40 empirical and 6 synthetic networks (see Appendix B for the  dataset details). For each network and IMP, we rank the eleven methods based on their performance. Hence, for each IMP, the overall performance of a method is defined as the average performance of the method over the 46 analyzed datasets.
For the eigenvalue minimization problem, we consider a measure $R_{\mu_1}$~\cite{schneider2011mitigation} to summarize the performance of a method in a given network:
\begin{equation}\label{eq:Rmeasure}
R_{\mu_1}=\frac{1}{Q\cdot\mu_1}\sum_{|\mathcal{S}|=1}^{Q}\mu_{1}'(\mathcal{S}),
\end{equation}
where $Q$ and $\mu_{1}'(\mathcal{S})$ denote the number of removed nodes and the largest eigenvalue after removing the nodes in $\mathcal{S}$, respectively, and $R_{\mu_1}$ represents the average largest eigenvalue within $q=|\mathcal{S}|/N\in[1/N,Q/N]$. In the calculation, we only considered a small fraction of removed nodes and set $Q=\lfloor0.2 N\rceil$.
Similarly, we define the average giant component  ($R_{G(\mathcal{S})}=\frac{1}{Q}\sum_{|\mathcal{S}|=1}^{Q}G(\mathcal{S})$), where $G(\mathcal{S})\in [0,1]$ denotes the relative size of the giant connected component after removing the nodes in $\mathcal{S}$~\cite{morone2015influence}. The average coverage for the linear threshold model is $R_{\sigma(\mathcal{S})}=\frac{1}{Q}\sum_{|\mathcal{S}|=1}^{Q}\sigma(\mathcal{S})$, where $\sigma(\mathcal{S})$ denotes the fraction of activated nodes in the processes initiated by the nodes in $\mathcal{S}$.

\begin{figure*}
  \centering
  \includegraphics[width=5in]{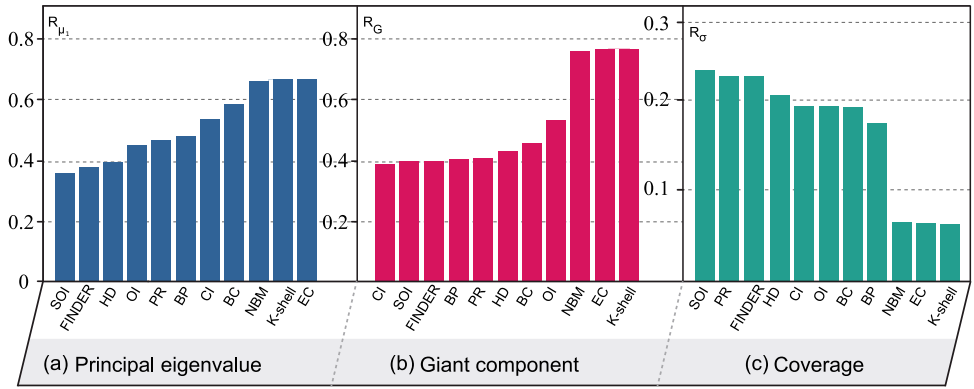}\\
  \caption{\textbf{Methods' average performance.} Average $R_{\mu_1}$, $R_{G}$, and $R_{\sigma}$ for the analyzed methods.
  The SOI method achieves the best performance in terms of $R_{\mu_1}$ and $R_{\sigma}$, and near-best performance in $R_{G}$.}
  \label{fig:rankingscore}
\end{figure*}

For each empirical network, we rank the eleven methods by $R_{\mu_1}$, $R_{G}$, and $R_{\sigma}$.
The overall performance of a method is defined as the average ranking score of the method across the 46 analyzed datasets. 
The SOI method achieves optimal or nearly optimal performance for the three evaluation metrics considered here, ($R_{\mu_1}$, $R_{G}$, and $R_{\sigma}$), as shown in Fig.~\ref{fig:rankingscore}.
Moreover, in the eigenvalue minimization problem, among the 40 analyzed real networks, the SOI method performs the best in 28 networks (70\%), whereas the CI method performs the best in only three networks (7.5\%). The SOI method achieves a considerably lower average $R_{\mu_1}$ and a better average ranking than the state-of-the-art methods (Fig.~3a and Appendix Fig. \ref{fig:ranking}).
In the eigenvalue minimization problem, the optimal performance of the SOI method is reasonable, given that the method was introduced to solve this specific problem. However, we show that the performance of the SOI method is also nearly optimal or optimal in the giant component and spreading coverage problems, indicating better generalization properties associated with the SOI method than with other methods.

In particular, in the giant component problem, among the 40 analyzed real networks, the SOI and CI methods performed the best in 14 (35\%) and 5 (20\%) networks, respectively. Surprisingly, even though the SOI was not specifically designed to solve the giant component problem, it is the best-performing method in a larger number of networks than the current state-of-the-art method (CI).
According to the average $R_{G}$, the CI is the best-performing method, as expected from previous studies ~\cite{morone2015influence}. However, the SOI exhibits a comparable performance, being outperformed by the CI method by less than 2\%.
 The other methods perform substantially worse than CI and SOI.

For the spreading maximization problem, over the analyzed 40 real networks, the SOI
performed the best in 24 (60\%)
networks. 
Other popular metrics for the identification of influential spreaders, such as the K-shell, NBM, and EC, exhibit worse performance than SOI. The main reason is that the three methods (K-shell, NBM, and EC) were aimed at characterizing the importance of a single node rather than multiple nodes. As $|\mathcal{S}|$ increases, the OI mechanism plays a significant role in the set influence, which degrades the performance of the K-shell, NBM, and EC methods.

In addition to the above IMPs, we also studied the influence blocking problem for two epidemic spreading models in the supercritical regime: the SIS and SIR models.
The results are in qualitative agreement with those of the coverage maximization problem, in which the proposed method substantially outperform the existing methods (see Appendix C for all the details, and Appendix D for the results).

\begin{figure}
  \centering
  \includegraphics[trim={0cm 0.2cm 0cm 0cm},clip=true,width=3in]{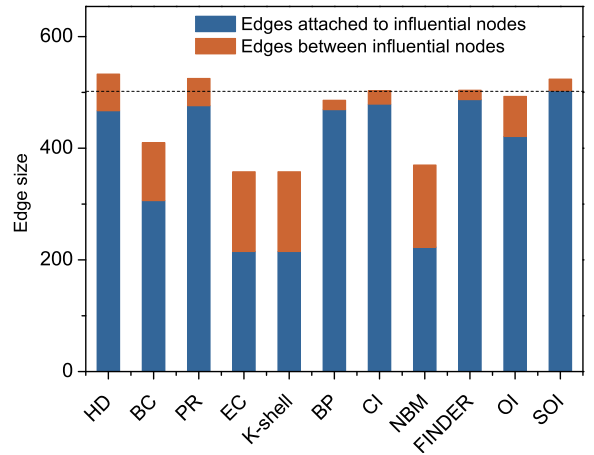}\\
  \caption{\textbf{Role of the set and OIs.} We compared the number of edges attached to the detected nodes (set influence, blue bars' height) and the number of ``intra-set" edges between the detected nodes (OI, orange bars' height). We select 10\% nodes as the eigenshield nodes in the Euroroads network.
 The SOI method exhibits the smallest number of intra-set edges, whereas a large fraction of intra-set edges are observed for other methods. 
  }
  \label{fig:sumandbetweenedge}
\end{figure}

To better understand the superior performance of the SOI method over state-of-the-art methods, we compare the cumulative degree of the detected nodes and the internal connectedness between the detected nodes in the Euroroads network (see Fig. \ref{fig:sumandbetweenedge}). Internal connectedness is represented by the OI between detected nodes. The simplified set influence $w(\mathcal{S})$ consists of two parts: the cumulative degree of nodes (the first part of the right-hand side (r.h.s.) of Eq. \eqref{eq:soiformalism}, $k_{sum}=\sum_{i\in\mathcal{S}}k_i$) and the edges between eigenshield nodes (defined by the second part of the r.h.s. of Eq. \eqref{eq:soiformalism}, $k_{int}=\frac{1}{2}\sum_{i\in\mathcal{S}} k_i^{\text{(int)}}(\mathcal{S})$).
Since $w(\mathcal{S})=k_{sum}-k_{int}$, to maximize $w(\mathcal{S})$, we should maximize $k_{sum}$ and simultaneously minimize $k_{int}$.
In Fig. \ref{fig:sumandbetweenedge}, the SOI method maximizes the $w(\mathcal{S})$ in Eq. \eqref{eq:soiformalism}, having the largest cumulative degree of nodes, but low internal connectedness. In contrast, the state-of-the-art methods only consider the node influence and exhibit considerable OIs between nodes. Therefore, they cannot achieve the maximum $w(\mathcal{S})$. ( See Appendix Fig. \ref{fig:overlaprealdegree} for the similar results in other networks.)

\section{Conclusion}

To summarize, 
we develop a theoretical framework to detect the eigenshield of a network, namely, the small set of nodes that, when removed, causes the largest drop in the network's spectral radius.
The identification of the eigenshield nodes is rephrased as the problem of optimizing the set influence function, which is solved through a greedy algorithm scaled for large complex networks. The proposed method could remove insignificant nodes from the eigenshield nodes, in contrast with classical methods that only add new nodes to  maximize the influence of eigenshield node set.
 The method proposed herein not only exhibits optimal or nearly optimal performance in diverse influence maximization problems compared to the state-of-the-art methods, but the influence redundancy also explains the drawbacks of the classical methods. Although differences in the specifications of network dynamics can affect the performance of the methods~\cite{de2014role,aral2018social,iannelli2018influencers,zhou2019fast,mariani2020network}, our findings point toward a method to identify eigenshield nodes of complex networks with better generalization properties than existing methods. 


While our study focused on monopartite networks, many real-world systems (such as social and transportation networks) can be alternatively described as networks of interacting networks~\cite{dong2021optimal}, higher-order effects~\cite{battiston2020networks}, and temporal effects~\cite{liao2017ranking}.
Future research may generalize the proposed method to more complex representations, which will require the development of matrix perturbation techniques for appropriate matrix representations of the interactions (e.g., adjacency tensors).




\section{Materials and Methods}

\subsection{Network perturbation}
We parameterize the perturbation of a network as $\mathsf{A}'=\mathsf{A}+\varepsilon \mathsf{P}$, where $\varepsilon \mathsf{P}$ denotes an arbitrary perturbation matrix. The terms $\{\mu_i\}$ and $\{\mathbf{v}_i\}$ denote the eigenvalues and corresponding eigenvectors of $\mathsf{A}$ (with $\mu_1\geq \mu_2\geq...\geq \mu_N$, $||\mathbf{v}_i||_2=1$). Similarly, $\{\mu'_i\}$ and $\{\mathbf{v}'_i\}$ denote the eigenvalues and corresponding eigenvectors of $\mathsf{A}'$.
 According to the algebraic theory, 
 the eigenvalues $\{\mu'_i\}$ of the perturbed matrix, $\mathsf{A}'$, can be expressed by the Taylor series as
 \begin{equation}
\mu_i'=\mu'_i(\varepsilon,P)=\mu_i+\sum_{j=1}^{+\infty} k_{i,j}\varepsilon^j,
\end{equation}
where $k_{i,j}$ represents the $j$-th order Taylor coefficient for eigenvalue $\mu'_i$.
We can derive an analytical expression for the first-order Taylor coefficient~\cite{restrepo2006characterizing}
\begin{equation}\label{eq:basic}
k_{i,1}=\mathbf{v}_i^TP\mathbf{v}_i,
\end{equation}
where $||\mathbf{v}_i||_2=1$ (See Appendix E for the derivation). Equation \eqref{eq:basic} forms the basis of our method for identifying a set of eigenshield nodes.

\subsection{Eigenvalue-based influence of nodes}
We treat the node removal as network perturbation. The perturbation matrix $\mathsf{R}=(p_{ij})_{N\times N}$ should be $r_{ij}=r_{ji}=a_{ij}$ if the endpoints $i\in \mathcal{S}$ or $j\in \mathcal{S}$; $r_{ij}=0$ otherwise. Based on Eq. \ref{eq:basic}, the eigenvalues of the remaining network's adjacency matrix $\mathsf{A}'=\mathsf{A}+\epsilon \mathsf{R} \quad (\epsilon=-1)$ are
\begin{equation}
\label{eq:eigenvaluenetperturb5}
\mu_i(\epsilon=-1)\approx
\mu_i-
\sum_{(a,b)\in\mathcal{E}_{\mathcal{S}}}v_{ia}v_{ib},
\end{equation}
where $|\mathbf{v}_i|=1$.
Recalling Eqs. \ref{eq:decrease}$\sim$\ref{eq:setandoverlapinfluence}, the influence overlap among the eigenshield nodes is the sum of individual influence minus the set influence,



\begin{equation}
\label{eq:overlap}
\Delta \mu_{i,{ov}}(\mathcal{S})=
\sum_{a\in{\mathcal{S}}}\Delta \mu_i(a)-\Delta \mu_{i}(\mathcal{S}).
\end{equation}
Particularly, we investigate the influence overlap between two eigenshield nodes,
\begin{equation}
\label{eq:overlap1}
\Delta \mu_{i,{ov}}(a,b)=
\Delta \mu_i(a)+\Delta \mu_i(b)-\Delta \mu_{i}(\{a,b\})=s_{iab},
\end{equation}
which actually means the influence of the common edge between two nodes $a$ and $b$.
Furthermore, the influence overlap among nodes in a set is
\begin{equation}
\label{eq:overlap3}
\Delta\mu_{i,\text{ov}}(\mathcal{S})=  \sum_{(a,b)\in\mathcal{E}^{\text{int}}_{\mathcal{S}}}s_{iab}=\sum_{(a,b)\in\mathcal{E}^{\text{int}}_{\mathcal{S}}}\Delta \mu_{i,{ov}}(a,b).
\end{equation}

When minimizing the largest eigenvalue of $A'$, we should consider all $\mu_i,i=1,2,..,N$ (Eq. \ref{eq:setinfluence}) that is of high time complexity.
Note that the $ \tilde{\lambda}_1$ is more likely to be determined by the decrease of large $\mu_i$ than small $\mu_i$, we introduce the weighted sum of $\Delta \mu_{i}(\mathcal{S})$,
\begin{equation}
\label{eq:sum1}
w(\mathcal{S}) =\sum_i \mu_i\Delta\mu_i(\mathcal{S}),
\end{equation}
Hence, the problem of minimizing $\tilde{\lambda}_1$ is rephrased of maximizing $w(\mathcal{S})$ in Eq. \ref{eq:sum1}.

For the undirected networks, supposing that there are no multiplicity eigenvalues, the symmetric matrix could be decomposed into
\begin{equation}
\label{eq:sum2}
\mathsf{A}=[\mathbf{v}_1,\mathbf{v}_2,...,\mathbf{v}_N]
\begin{pmatrix}
                                                                                                    \mu_1 & 0 & \vdots & 0 \\
                                                                                                    0 & \mu_2 & \vdots & 0 \\
                                                                                                    ...& ... & \ddots & ... \\
                                                                                                    0 & 0 & \vdots & \mu_N \\
                                                                                                  \end{pmatrix}
                                                                                                  [\mathbf{v}_1,\mathbf{v}_2,...,\mathbf{v}_N]^T
.
\end{equation}
Equation \ref{eq:sum2} actually represents the eigenvector decomposition, from which we get $\sum_i\mu_i s_{iab}=\sum_i\mu_iv_{ia}v_{ib}=1$ if $(a,b)\in E$.
Combining Eqs. \ref{eq:sum1} and \ref{eq:sum2}, we can abbreviate Eq. \ref{eq:sum1} into
\begin{equation}
\label{eq:sum4}
w(\mathcal{S}) =2(\sum_{i\in \mathcal{S}}k_i -\sum_{i\in \mathcal{S},j\in \mathcal{S},i<j}a_{ij}),
\end{equation}
where $k_i$ is the degree of node $i$. Neglecting the constant term, we arrive at Eq. \ref{eq:soiformalism} (we rewrite it here),
\begin{equation}
\label{eq:sum3}
 w(\mathcal{S})=\sum_{i\in\mathcal{S}}k_i-\frac{1}{2}\sum_{i\in\mathcal{S}} k_i^{\text{(int)}}(\mathcal{S}),
\end{equation}
$ w(\mathcal{S})$ means the sum  of edges attached to eigenshield nodes. Note that in Eq. \ref{eq:sum3}, the first term of the r.h.s of Eq. \ref{eq:sum3} is the cumulative sum degree of all eigenshield nodes, while the second term is the edges between eigenshield nodes that mean influence redundancy.

\subsection{Optimization algorithm}
We first directly optimize $\tilde{\lambda}_1$, and then optimize $w(\mathcal{S})$ to reduce time complexity.

\textbf{Optimizing $\tilde{\lambda}_1$:}
The algorithm of OI is as follows:

1. Initially, the eigenshield node set is empty, $\mathcal{S}=\emptyset$. The aim is to choose a fixed size of eigenshield nodes that minimize the largest eigenvalue of the remaining network. 

2. Calculate the  eigenvalues $\mu_i$ and the corresponding eigenvectors $\mathbf{v}_i$ for the adjacency matrix of a network, $|\mathbf{v}_i|=1$, $i=1,2,3,...h$.

3. Choose a new candidate node $a$ ($a\notin \mathcal{S}$) and add the node into $\mathcal{S}$. The newly chosen node should minimize $\tilde{\lambda}_1$. The decrease of $\tilde{\lambda}_1$ induced by choosing node $a$ is
\begin{equation}
\label{eq:contribution}
\Delta\tilde{\lambda}_1(a)=\tilde{\lambda}_1(\mathcal{S})-\tilde{\lambda}_1(\mathcal{S}\cup a).
\end{equation}

4. Re-calculate the contribution of previously chosen nodes to the $\tilde{\lambda}_1$. Because of the  influence overlap mechanism in Eq. \ref{eq:overlap}, the previously chosen nodes may have much  influence  overlap with newly chosen nodes. Thus, the previously important nodes may be insignificant after adding new nodes into $\mathcal{S}$. The contribution of a previous node $m$ (in $\mathcal{S}$) to the $\tilde{\lambda}_1$ is evaluated by
\begin{equation}
\label{eq:contribution2}
\Delta\tilde{\lambda}_1(m)=\tilde{\lambda}_1(\mathcal{S'}\setminus m)-\tilde{\lambda}_1(\mathcal{S'}),
\end{equation}
where $\mathcal{S'}=\mathcal{S}\cup a$ and $\mathcal{S'}\setminus m$ is the remaining set by removing node $m$ from $\mathcal{S'}$. Here, we choose the node with the smallest $\Delta\tilde{\lambda}_1(m)$, $m\neq a$. If $\Delta\tilde{\lambda}_1(m)<\Delta\tilde{\lambda}_1(a)$, we remove the previously chosen node $m$ from set $\mathcal{S}$; otherwise skip the step. This procedure actually means removing redundant nodes from $\mathcal{S'}$.

5. Repeat step 3--4 to choose the fix size of eigenshield nodes.

The proposed OI method is a greedy algorithm that minimizes Eq. \ref{eq:setinfluence}. Note that the key issue of the algorithm is step 3 and step 4 that simultaneously add marginal eigenshield nodes and remove existing chosen nodes, which reduces the influence overlap between eigenshield nodes. Consequently, the set influence could be maximized.

\textbf{Optimizing $w(\mathcal{S})$:}
The algorithm of SOI is as follows:

(1) Initially, the set of eigenshield nodes is empty, $\mathcal{S}=\emptyset$. The aim is to choose a fixed size of eigenshield nodes that maximize $w(\mathcal{S})$ in Eq. \ref{eq:sum3}.

2. At every step, we choose a new eigenshield node $a$ and add the node into $\mathcal{S}$. The newly chosen node should maximize $w(\mathcal{S})$ (Eq. \ref{eq:sum3}). The increase of $w(\mathcal{S})$ induced by choosing node $a$ is
\begin{equation}
\label{eq:contribution4}
w(a)=w(\mathcal{S}\cup a)-w(\mathcal{S})=\sum_{k\in V\setminus \mathcal{S}}a_{ak}.
\end{equation}
The newly node is actually the one with the largest degree after removing the node set $\mathcal{S}$.

3. Re-calculate the contribution of previously chosen nodes to the $w$. The contribution of a previous node $m$ (in $\mathcal{S}$) to the $w$ means the decrease of $w$ if we remove the node $m$ from $\mathcal{S}$, which is evaluated by
\begin{equation}
\label{eq:contribution5}
\Delta w(m)=w(\mathcal{S'})-w(\mathcal{S'}\setminus m)=\sum_{k\in V\setminus \mathcal{S'}}a_{mk}.
\end{equation}
where $\mathcal{S'}=\mathcal{S}\cup a$.
$\Delta w(m)$ actually means the size of connections between $m$ and the remaining network after removing the node set $\mathcal{S'}$.
Here, we choose the node with the smallest $\Delta w(m)$, $m\neq a$. If $\Delta w(m)<\Delta w(a)$, we remove the  node $m$ from set $\mathcal{S'}$; otherwise skip the step. This procedure actually means removing redundant nodes from $\mathcal{S'}$.

(4) Repeat step 2--3 to choose the fix number of eigenshield nodes.

In the proposed algorithms (OI and SOI),  we only use greedy strategy to optimize the objective function that might only reach local optimization. Other better intelligence algorithms may be introduced to optimize Eq. \ref{eq:sum3} (See the optimization methods in the supplementary of ref. \cite{morone2015influence}). In the experiments, the simple greedy algorithm still arrives at perfect performance.

\subsection{Analysis of the time complexity}
For the algorithm OI, we require calculating the eigenvalues and the corresponding eigenvectors that scales as $O(N^3)$. At each step, a candidate node is chosen by traversing the relevant eigenvectors that scale $O(h|E|)$, where $h$ is a small positive number in Eq. \ref{eq:setinfluence}. Besides, it needs at least $O(h|\mathcal{S}|)$ steps to scan all the chosen nodes to remove insignificant ones who contribute to the $\tilde{\lambda}_1$ less than the new node. Thus, the overall time complexity is $O(N^3+h|E|+h|\mathcal{S}|)=O(N^3)$, where the main time consumption is the calculation of eigenvalues and eigenvectors.

For the algorithm SOI, we do not require to calculate the eigenvalues and eigenvectors explicitly. At every step, we maximize the Eq. \ref{eq:sum3} that scales as $O(|E|)$. With the size increase of eigenshield nodes, the previous chosen nodes may be excluded from the eigenshield node set. The overall time complexity scales as $O(|E|\cdot|\mathcal{S}|)$. For sparse networks, $|E|\propto N$, the time complexity is $O(N\cdot|\mathcal{S}|)$.

\subsection{Data set and baseline methods}
We consider 40 real-world networks drawn from disparate fields, including infrastructure networks, social networks, protein-protein networks, biological networks, scientific collaboration networks, and so on. Besides, we also generate model networks, including scale free networks and ER networks. Please refer to Appendix B for the dataset details.

We compare our method with nine classical methods: high degree (HD), betweenness centrality (BC), PageRank (PR) index, eigenvector centrality (EC),
K-shell index, belief propagation index, collective
influence (CI), non-backtracking matrix (NBM), and FINDER index (a reinforcement learning method). Please refer to Appendix A for the
details of the state-of-the-art methods.

\begin{acknowledgments}
This work is jointly supported by the National Natural Science Foundation of China (11547040), and Tencent Open Research Fund. MSM acknowledges financial support from the URPP Social Networks at the University of Zurich.
\end{acknowledgments}

\section{Reference}

\begin{thebibliography}{42}%
\makeatletter
\providecommand \@ifxundefined [1]{%
 \@ifx{#1\undefined}
}%
\providecommand \@ifnum [1]{%
 \ifnum #1\expandafter \@firstoftwo
 \else \expandafter \@secondoftwo
 \fi
}%
\providecommand \@ifx [1]{%
 \ifx #1\expandafter \@firstoftwo
 \else \expandafter \@secondoftwo
 \fi
}%
\providecommand \natexlab [1]{#1}%
\providecommand \enquote  [1]{``#1''}%
\providecommand \bibnamefont  [1]{#1}%
\providecommand \bibfnamefont [1]{#1}%
\providecommand \citenamefont [1]{#1}%
\providecommand \href@noop [0]{\@secondoftwo}%
\providecommand \href [0]{\begingroup \@sanitize@url \@href}%
\providecommand \@href[1]{\@@startlink{#1}\@@href}%
\providecommand \@@href[1]{\endgroup#1\@@endlink}%
\providecommand \@sanitize@url [0]{\catcode `\\12\catcode `\$12\catcode
  `\&12\catcode `\#12\catcode `\^12\catcode `\_12\catcode `\%12\relax}%
\providecommand \@@startlink[1]{}%
\providecommand \@@endlink[0]{}%
\providecommand \url  [0]{\begingroup\@sanitize@url \@url }%
\providecommand \@url [1]{\endgroup\@href {#1}{\urlprefix }}%
\providecommand \urlprefix  [0]{URL }%
\providecommand \Eprint [0]{\href }%
\providecommand \doibase [0]{http://dx.doi.org/}%
\providecommand \selectlanguage [0]{\@gobble}%
\providecommand \bibinfo  [0]{\@secondoftwo}%
\providecommand \bibfield  [0]{\@secondoftwo}%
\providecommand \translation [1]{[#1]}%
\providecommand \BibitemOpen [0]{}%
\providecommand \bibitemStop [0]{}%
\providecommand \bibitemNoStop [0]{.\EOS\space}%
\providecommand \EOS [0]{\spacefactor3000\relax}%
\providecommand \BibitemShut  [1]{\csname bibitem#1\endcsname}%
\let\auto@bib@innerbib\@empty
\bibitem [{\citenamefont {Morone}\ \emph {et~al.}(2015)\citenamefont {Morone},
  \citenamefont {Makse} \emph {et~al.}}]{morone2015influence}%
  \BibitemOpen
  \bibfield  {author} {\bibinfo {author} {\bibfnamefont {F.}~\bibnamefont
  {Morone}}, \bibinfo {author} {\bibfnamefont {H.}~\bibnamefont {Makse}},
  \emph {et~al.},\ }\href@noop {} {\bibfield  {journal} {\bibinfo  {journal}
  {Nature}\ }\textbf {\bibinfo {volume} {524}},\ \bibinfo {pages} {65}
  (\bibinfo {year} {2015})}\BibitemShut {NoStop}%
\bibitem [{\citenamefont {L{\"u}}\ \emph {et~al.}(2016)\citenamefont {L{\"u}},
  \citenamefont {Chen}, \citenamefont {Ren}, \citenamefont {Zhang},
  \citenamefont {Zhang},\ and\ \citenamefont {Zhou}}]{lu2016vital}%
  \BibitemOpen
  \bibfield  {author} {\bibinfo {author} {\bibfnamefont {L.}~\bibnamefont
  {L{\"u}}}, \bibinfo {author} {\bibfnamefont {D.}~\bibnamefont {Chen}},
  \bibinfo {author} {\bibfnamefont {X.-L.}\ \bibnamefont {Ren}}, \bibinfo
  {author} {\bibfnamefont {Q.-M.}\ \bibnamefont {Zhang}}, \bibinfo {author}
  {\bibfnamefont {Y.-C.}\ \bibnamefont {Zhang}}, \ and\ \bibinfo {author}
  {\bibfnamefont {T.}~\bibnamefont {Zhou}},\ }\href@noop {} {\bibfield
  {journal} {\bibinfo  {journal} {Physics Reports}\ }\textbf {\bibinfo {volume}
  {650}},\ \bibinfo {pages} {1} (\bibinfo {year} {2016})}\BibitemShut {NoStop}%
\bibitem [{\citenamefont {Pei}\ \emph {et~al.}(2020)\citenamefont {Pei},
  \citenamefont {Wang}, \citenamefont {Morone},\ and\ \citenamefont
  {Makse}}]{pei2020influencer}%
  \BibitemOpen
  \bibfield  {author} {\bibinfo {author} {\bibfnamefont {S.}~\bibnamefont
  {Pei}}, \bibinfo {author} {\bibfnamefont {J.}~\bibnamefont {Wang}}, \bibinfo
  {author} {\bibfnamefont {F.}~\bibnamefont {Morone}}, \ and\ \bibinfo {author}
  {\bibfnamefont {H.~A.}\ \bibnamefont {Makse}},\ }\href@noop {} {\bibfield
  {journal} {\bibinfo  {journal} {Journal of Complex Networks}\ }\textbf
  {\bibinfo {volume} {8}},\ \bibinfo {pages} {cnz029} (\bibinfo {year}
  {2020})}\BibitemShut {NoStop}%
\bibitem [{\citenamefont {Wang}\ \emph {et~al.}(2016)\citenamefont {Wang},
  \citenamefont {Bauch}, \citenamefont {Bhattacharyya}, \citenamefont
  {d'Onofrio}, \citenamefont {Manfredi}, \citenamefont {Perc}, \citenamefont
  {Perra}, \citenamefont {Salath{\'e}},\ and\ \citenamefont
  {Zhao}}]{wang2016statistical}%
  \BibitemOpen
  \bibfield  {author} {\bibinfo {author} {\bibfnamefont {Z.}~\bibnamefont
  {Wang}}, \bibinfo {author} {\bibfnamefont {C.~T.}\ \bibnamefont {Bauch}},
  \bibinfo {author} {\bibfnamefont {S.}~\bibnamefont {Bhattacharyya}}, \bibinfo
  {author} {\bibfnamefont {A.}~\bibnamefont {d'Onofrio}}, \bibinfo {author}
  {\bibfnamefont {P.}~\bibnamefont {Manfredi}}, \bibinfo {author}
  {\bibfnamefont {M.}~\bibnamefont {Perc}}, \bibinfo {author} {\bibfnamefont
  {N.}~\bibnamefont {Perra}}, \bibinfo {author} {\bibfnamefont
  {M.}~\bibnamefont {Salath{\'e}}}, \ and\ \bibinfo {author} {\bibfnamefont
  {D.}~\bibnamefont {Zhao}},\ }\href@noop {} {\bibfield  {journal} {\bibinfo
  {journal} {Physics Reports}\ }\textbf {\bibinfo {volume} {664}},\ \bibinfo
  {pages} {1} (\bibinfo {year} {2016})}\BibitemShut {NoStop}%
\bibitem [{\citenamefont {Wang}\ and\ \citenamefont
  {Rong}(2009)}]{wang2009cascade}%
  \BibitemOpen
  \bibfield  {author} {\bibinfo {author} {\bibfnamefont {J.-W.}\ \bibnamefont
  {Wang}}\ and\ \bibinfo {author} {\bibfnamefont {L.-L.}\ \bibnamefont
  {Rong}},\ }\href@noop {} {\bibfield  {journal} {\bibinfo  {journal} {Safety
  science}\ }\textbf {\bibinfo {volume} {47}},\ \bibinfo {pages} {1332}
  (\bibinfo {year} {2009})}\BibitemShut {NoStop}%
\bibitem [{\citenamefont {Saavedra}\ \emph {et~al.}(2011)\citenamefont
  {Saavedra}, \citenamefont {Stouffer}, \citenamefont {Uzzi},\ and\
  \citenamefont {Bascompte}}]{saavedra2011strong}%
  \BibitemOpen
  \bibfield  {author} {\bibinfo {author} {\bibfnamefont {S.}~\bibnamefont
  {Saavedra}}, \bibinfo {author} {\bibfnamefont {D.~B.}\ \bibnamefont
  {Stouffer}}, \bibinfo {author} {\bibfnamefont {B.}~\bibnamefont {Uzzi}}, \
  and\ \bibinfo {author} {\bibfnamefont {J.}~\bibnamefont {Bascompte}},\
  }\href@noop {} {\bibfield  {journal} {\bibinfo  {journal} {Nature}\ }\textbf
  {\bibinfo {volume} {478}},\ \bibinfo {pages} {233} (\bibinfo {year}
  {2011})}\BibitemShut {NoStop}%
\bibitem [{\citenamefont {Dom{\'\i}nguez-Garc{\'\i}a}\ and\ \citenamefont
  {Munoz}(2015)}]{dominguez2015ranking}%
  \BibitemOpen
  \bibfield  {author} {\bibinfo {author} {\bibfnamefont {V.}~\bibnamefont
  {Dom{\'\i}nguez-Garc{\'\i}a}}\ and\ \bibinfo {author} {\bibfnamefont {M.~A.}\
  \bibnamefont {Munoz}},\ }\href@noop {} {\bibfield  {journal} {\bibinfo
  {journal} {Scientific reports}\ }\textbf {\bibinfo {volume} {5}},\ \bibinfo
  {pages} {1} (\bibinfo {year} {2015})}\BibitemShut {NoStop}%
\bibitem [{\citenamefont {Kempe}\ \emph {et~al.}(2003)\citenamefont {Kempe},
  \citenamefont {Kleinberg},\ and\ \citenamefont
  {Tardos}}]{kempe2003maximizing}%
  \BibitemOpen
  \bibfield  {author} {\bibinfo {author} {\bibfnamefont {D.}~\bibnamefont
  {Kempe}}, \bibinfo {author} {\bibfnamefont {J.}~\bibnamefont {Kleinberg}}, \
  and\ \bibinfo {author} {\bibfnamefont {{\'E}.}~\bibnamefont {Tardos}},\ }in\
  \href@noop {} {\emph {\bibinfo {booktitle} {The 9th ACM SIGKDD}}}\ (\bibinfo
  {organization} {ACM},\ \bibinfo {year} {2003})\ pp.\ \bibinfo {pages}
  {137--146}\BibitemShut {NoStop}%
\bibitem [{\citenamefont {Kitsak}\ \emph {et~al.}(2010)\citenamefont {Kitsak},
  \citenamefont {Gallos}, \citenamefont {Havlin}, \citenamefont {Liljeros},
  \citenamefont {Muchnik}, \citenamefont {Stanley},\ and\ \citenamefont
  {Makse}}]{kitsak2010identification}%
  \BibitemOpen
  \bibfield  {author} {\bibinfo {author} {\bibfnamefont {M.}~\bibnamefont
  {Kitsak}}, \bibinfo {author} {\bibfnamefont {L.~K.}\ \bibnamefont {Gallos}},
  \bibinfo {author} {\bibfnamefont {S.}~\bibnamefont {Havlin}}, \bibinfo
  {author} {\bibfnamefont {F.}~\bibnamefont {Liljeros}}, \bibinfo {author}
  {\bibfnamefont {L.}~\bibnamefont {Muchnik}}, \bibinfo {author} {\bibfnamefont
  {H.~E.}\ \bibnamefont {Stanley}}, \ and\ \bibinfo {author} {\bibfnamefont
  {H.~A.}\ \bibnamefont {Makse}},\ }\href@noop {} {\bibfield  {journal}
  {\bibinfo  {journal} {Nature physics}\ }\textbf {\bibinfo {volume} {6}},\
  \bibinfo {pages} {888} (\bibinfo {year} {2010})}\BibitemShut {NoStop}%
\bibitem [{\citenamefont {Wong}\ and\ \citenamefont
  {Collins}(2020)}]{wong2020evidence}%
  \BibitemOpen
  \bibfield  {author} {\bibinfo {author} {\bibfnamefont {F.}~\bibnamefont
  {Wong}}\ and\ \bibinfo {author} {\bibfnamefont {J.~J.}\ \bibnamefont
  {Collins}},\ }\href@noop {} {\bibfield  {journal} {\bibinfo  {journal}
  {Proceedings of the National Academy of Sciences}\ }\textbf {\bibinfo
  {volume} {117}},\ \bibinfo {pages} {29416} (\bibinfo {year}
  {2020})}\BibitemShut {NoStop}%
\bibitem [{\citenamefont {De~Arruda}\ \emph {et~al.}(2014)\citenamefont
  {De~Arruda}, \citenamefont {Barbieri}, \citenamefont {Rodr{\'\i}guez},
  \citenamefont {Rodrigues}, \citenamefont {Moreno},\ and\ \citenamefont
  {da~Fontoura~Costa}}]{de2014role}%
  \BibitemOpen
  \bibfield  {author} {\bibinfo {author} {\bibfnamefont {G.~F.}\ \bibnamefont
  {De~Arruda}}, \bibinfo {author} {\bibfnamefont {A.~L.}\ \bibnamefont
  {Barbieri}}, \bibinfo {author} {\bibfnamefont {P.~M.}\ \bibnamefont
  {Rodr{\'\i}guez}}, \bibinfo {author} {\bibfnamefont {F.~A.}\ \bibnamefont
  {Rodrigues}}, \bibinfo {author} {\bibfnamefont {Y.}~\bibnamefont {Moreno}}, \
  and\ \bibinfo {author} {\bibfnamefont {L.}~\bibnamefont
  {da~Fontoura~Costa}},\ }\href@noop {} {\bibfield  {journal} {\bibinfo
  {journal} {Physical Review E}\ }\textbf {\bibinfo {volume} {90}},\ \bibinfo
  {pages} {032812} (\bibinfo {year} {2014})}\BibitemShut {NoStop}%
\bibitem [{\citenamefont {Radicchi}\ and\ \citenamefont
  {Castellano}(2017)}]{radicchi2017fundamental}%
  \BibitemOpen
  \bibfield  {author} {\bibinfo {author} {\bibfnamefont {F.}~\bibnamefont
  {Radicchi}}\ and\ \bibinfo {author} {\bibfnamefont {C.}~\bibnamefont
  {Castellano}},\ }\href@noop {} {\bibfield  {journal} {\bibinfo  {journal}
  {Physical Review E}\ }\textbf {\bibinfo {volume} {95}},\ \bibinfo {pages}
  {012318} (\bibinfo {year} {2017})}\BibitemShut {NoStop}%
\bibitem [{\citenamefont {Iannelli}\ \emph {et~al.}(2018)\citenamefont
  {Iannelli}, \citenamefont {Mariani},\ and\ \citenamefont
  {Sokolov}}]{iannelli2018influencers}%
  \BibitemOpen
  \bibfield  {author} {\bibinfo {author} {\bibfnamefont {F.}~\bibnamefont
  {Iannelli}}, \bibinfo {author} {\bibfnamefont {M.~S.}\ \bibnamefont
  {Mariani}}, \ and\ \bibinfo {author} {\bibfnamefont {I.~M.}\ \bibnamefont
  {Sokolov}},\ }\href@noop {} {\bibfield  {journal} {\bibinfo  {journal}
  {Physical Review E}\ }\textbf {\bibinfo {volume} {98}},\ \bibinfo {pages}
  {062302} (\bibinfo {year} {2018})}\BibitemShut {NoStop}%
\bibitem [{\citenamefont {Aral}\ and\ \citenamefont
  {Dhillon}(2018)}]{aral2018social}%
  \BibitemOpen
  \bibfield  {author} {\bibinfo {author} {\bibfnamefont {S.}~\bibnamefont
  {Aral}}\ and\ \bibinfo {author} {\bibfnamefont {P.~S.}\ \bibnamefont
  {Dhillon}},\ }\href@noop {} {\bibfield  {journal} {\bibinfo  {journal}
  {Nature human behaviour}\ }\textbf {\bibinfo {volume} {2}},\ \bibinfo {pages}
  {375} (\bibinfo {year} {2018})}\BibitemShut {NoStop}%
\bibitem [{\citenamefont {Mariani}\ and\ \citenamefont
  {L{\"u}}(2020)}]{mariani2020network}%
  \BibitemOpen
  \bibfield  {author} {\bibinfo {author} {\bibfnamefont {M.~S.}\ \bibnamefont
  {Mariani}}\ and\ \bibinfo {author} {\bibfnamefont {L.}~\bibnamefont
  {L{\"u}}},\ }\href@noop {} {\bibfield  {journal} {\bibinfo  {journal}
  {Journal of Physics: Complexity}\ }\textbf {\bibinfo {volume} {1}},\ \bibinfo
  {pages} {011001} (\bibinfo {year} {2020})}\BibitemShut {NoStop}%
\bibitem [{\citenamefont {Ji}\ \emph {et~al.}(2017)\citenamefont {Ji},
  \citenamefont {L{\"u}}, \citenamefont {Yeung},\ and\ \citenamefont
  {Hu}}]{ji2017effective}%
  \BibitemOpen
  \bibfield  {author} {\bibinfo {author} {\bibfnamefont {S.}~\bibnamefont
  {Ji}}, \bibinfo {author} {\bibfnamefont {L.}~\bibnamefont {L{\"u}}}, \bibinfo
  {author} {\bibfnamefont {C.~H.}\ \bibnamefont {Yeung}}, \ and\ \bibinfo
  {author} {\bibfnamefont {Y.}~\bibnamefont {Hu}},\ }\href@noop {} {\bibfield
  {journal} {\bibinfo  {journal} {New Journal of Physics}\ }\textbf {\bibinfo
  {volume} {19}},\ \bibinfo {pages} {073020} (\bibinfo {year}
  {2017})}\BibitemShut {NoStop}%
\bibitem [{\citenamefont {Chakrabarti}\ \emph {et~al.}(2008)\citenamefont
  {Chakrabarti}, \citenamefont {Wang}, \citenamefont {Wang}, \citenamefont
  {Leskovec},\ and\ \citenamefont {Faloutsos}}]{chakrabarti2008epidemic}%
  \BibitemOpen
  \bibfield  {author} {\bibinfo {author} {\bibfnamefont {D.}~\bibnamefont
  {Chakrabarti}}, \bibinfo {author} {\bibfnamefont {Y.}~\bibnamefont {Wang}},
  \bibinfo {author} {\bibfnamefont {C.}~\bibnamefont {Wang}}, \bibinfo {author}
  {\bibfnamefont {J.}~\bibnamefont {Leskovec}}, \ and\ \bibinfo {author}
  {\bibfnamefont {C.}~\bibnamefont {Faloutsos}},\ }\href@noop {} {\bibfield
  {journal} {\bibinfo  {journal} {ACM Transactions on Information and System
  Security (TISSEC)}\ }\textbf {\bibinfo {volume} {10}},\ \bibinfo {pages} {1}
  (\bibinfo {year} {2008})}\BibitemShut {NoStop}%
\bibitem [{\citenamefont {Staniczenko}\ \emph {et~al.}(2013)\citenamefont
  {Staniczenko}, \citenamefont {Kopp},\ and\ \citenamefont
  {Allesina}}]{staniczenko2013ghost}%
  \BibitemOpen
  \bibfield  {author} {\bibinfo {author} {\bibfnamefont {P.~P.}\ \bibnamefont
  {Staniczenko}}, \bibinfo {author} {\bibfnamefont {J.~C.}\ \bibnamefont
  {Kopp}}, \ and\ \bibinfo {author} {\bibfnamefont {S.}~\bibnamefont
  {Allesina}},\ }\href@noop {} {\bibfield  {journal} {\bibinfo  {journal}
  {Nature communications}\ }\textbf {\bibinfo {volume} {4}},\ \bibinfo {pages}
  {1} (\bibinfo {year} {2013})}\BibitemShut {NoStop}%
\bibitem [{\citenamefont {Restrepo}\ \emph {et~al.}(2006)\citenamefont
  {Restrepo}, \citenamefont {Ott},\ and\ \citenamefont
  {Hunt}}]{restrepo2006characterizing}%
  \BibitemOpen
  \bibfield  {author} {\bibinfo {author} {\bibfnamefont {J.~G.}\ \bibnamefont
  {Restrepo}}, \bibinfo {author} {\bibfnamefont {E.}~\bibnamefont {Ott}}, \
  and\ \bibinfo {author} {\bibfnamefont {B.~R.}\ \bibnamefont {Hunt}},\
  }\href@noop {} {\bibfield  {journal} {\bibinfo  {journal} {Physical review
  letters}\ }\textbf {\bibinfo {volume} {97}},\ \bibinfo {pages} {094102}
  (\bibinfo {year} {2006})}\BibitemShut {NoStop}%
\bibitem [{\citenamefont {Albert}\ \emph {et~al.}(2004)\citenamefont {Albert},
  \citenamefont {Albert},\ and\ \citenamefont
  {Nakarado}}]{albert2004structural}%
  \BibitemOpen
  \bibfield  {author} {\bibinfo {author} {\bibfnamefont {R.}~\bibnamefont
  {Albert}}, \bibinfo {author} {\bibfnamefont {I.}~\bibnamefont {Albert}}, \
  and\ \bibinfo {author} {\bibfnamefont {G.~L.}\ \bibnamefont {Nakarado}},\
  }\href@noop {} {\bibfield  {journal} {\bibinfo  {journal} {Physical review
  E}\ }\textbf {\bibinfo {volume} {69}},\ \bibinfo {pages} {025103} (\bibinfo
  {year} {2004})}\BibitemShut {NoStop}%
\bibitem [{\citenamefont {Krzakala}\ \emph {et~al.}(2013)\citenamefont
  {Krzakala}, \citenamefont {Moore}, \citenamefont {Mossel}, \citenamefont
  {Neeman}, \citenamefont {Sly}, \citenamefont {Zdeborová},\ and\
  \citenamefont {Zhang}}]{krzakala2013spectral}%
  \BibitemOpen
  \bibfield  {author} {\bibinfo {author} {\bibfnamefont {F.}~\bibnamefont
  {Krzakala}}, \bibinfo {author} {\bibfnamefont {C.}~\bibnamefont {Moore}},
  \bibinfo {author} {\bibfnamefont {E.}~\bibnamefont {Mossel}}, \bibinfo
  {author} {\bibfnamefont {J.}~\bibnamefont {Neeman}}, \bibinfo {author}
  {\bibfnamefont {A.}~\bibnamefont {Sly}}, \bibinfo {author} {\bibfnamefont
  {L.}~\bibnamefont {Zdeborová}}, \ and\ \bibinfo {author} {\bibfnamefont
  {P.}~\bibnamefont {Zhang}},\ }\href@noop {} {\bibfield  {journal} {\bibinfo
  {journal} {Proceedings of the National Academy of Sciences}\ }\textbf
  {\bibinfo {volume} {110}},\ \bibinfo {pages} {20935} (\bibinfo {year}
  {2013})}\BibitemShut {NoStop}%
\bibitem [{\citenamefont {Mugisha}\ and\ \citenamefont
  {Zhou}(2016)}]{mugisha2016identifying}%
  \BibitemOpen
  \bibfield  {author} {\bibinfo {author} {\bibfnamefont {S.}~\bibnamefont
  {Mugisha}}\ and\ \bibinfo {author} {\bibfnamefont {H.-J.}\ \bibnamefont
  {Zhou}},\ }\href@noop {} {\bibfield  {journal} {\bibinfo  {journal} {Physical
  Review E}\ }\textbf {\bibinfo {volume} {94}},\ \bibinfo {pages} {012305}
  (\bibinfo {year} {2016})}\BibitemShut {NoStop}%
\bibitem [{\citenamefont {Page}\ \emph {et~al.}(1999)\citenamefont {Page},
  \citenamefont {Brin}, \citenamefont {Motwani},\ and\ \citenamefont
  {Winograd}}]{page1999pagerank}%
  \BibitemOpen
  \bibfield  {author} {\bibinfo {author} {\bibfnamefont {L.}~\bibnamefont
  {Page}}, \bibinfo {author} {\bibfnamefont {S.}~\bibnamefont {Brin}}, \bibinfo
  {author} {\bibfnamefont {R.}~\bibnamefont {Motwani}}, \ and\ \bibinfo
  {author} {\bibfnamefont {T.}~\bibnamefont {Winograd}},\ }\href@noop {} {\emph
  {\bibinfo {title} {The PageRank citation ranking: Bringing order to the
  web}}},\ \bibinfo {type} {Tech. Rep.}\ (\bibinfo  {institution} {Stanford
  InfoLab},\ \bibinfo {year} {1999})\BibitemShut {NoStop}%
\bibitem [{\citenamefont {Ren}\ \emph {et~al.}(2019)\citenamefont {Ren},
  \citenamefont {Gleinig}, \citenamefont {Helbing},\ and\ \citenamefont
  {Antulov-Fantulin}}]{ren2019generalized}%
  \BibitemOpen
  \bibfield  {author} {\bibinfo {author} {\bibfnamefont {X.-L.}\ \bibnamefont
  {Ren}}, \bibinfo {author} {\bibfnamefont {N.}~\bibnamefont {Gleinig}},
  \bibinfo {author} {\bibfnamefont {D.}~\bibnamefont {Helbing}}, \ and\
  \bibinfo {author} {\bibfnamefont {N.}~\bibnamefont {Antulov-Fantulin}},\
  }\href@noop {} {\bibfield  {journal} {\bibinfo  {journal} {Proceedings of the
  national academy of sciences}\ }\textbf {\bibinfo {volume} {116}},\ \bibinfo
  {pages} {6554} (\bibinfo {year} {2019})}\BibitemShut {NoStop}%
\bibitem [{\citenamefont {Brin}\ and\ \citenamefont
  {Page}(1998)}]{brin1998anatomy}%
  \BibitemOpen
  \bibfield  {author} {\bibinfo {author} {\bibfnamefont {S.}~\bibnamefont
  {Brin}}\ and\ \bibinfo {author} {\bibfnamefont {L.}~\bibnamefont {Page}},\
  }\href@noop {} {\bibfield  {journal} {\bibinfo  {journal} {Computer networks
  and ISDN systems}\ }\textbf {\bibinfo {volume} {30}},\ \bibinfo {pages} {107}
  (\bibinfo {year} {1998})}\BibitemShut {NoStop}%
\bibitem [{\citenamefont {Fan}\ \emph {et~al.}(2020)\citenamefont {Fan},
  \citenamefont {Zeng}, \citenamefont {Sun},\ and\ \citenamefont
  {Liu}}]{fan2020finding}%
  \BibitemOpen
  \bibfield  {author} {\bibinfo {author} {\bibfnamefont {C.}~\bibnamefont
  {Fan}}, \bibinfo {author} {\bibfnamefont {L.}~\bibnamefont {Zeng}}, \bibinfo
  {author} {\bibfnamefont {Y.}~\bibnamefont {Sun}}, \ and\ \bibinfo {author}
  {\bibfnamefont {Y.-Y.}\ \bibnamefont {Liu}},\ }\href@noop {} {\bibfield
  {journal} {\bibinfo  {journal} {Nature Machine Intelligence}\ }\textbf
  {\bibinfo {volume} {2}},\ \bibinfo {pages} {317} (\bibinfo {year}
  {2020})}\BibitemShut {NoStop}%
\bibitem [{\citenamefont {Schneider}\ \emph {et~al.}(2011)\citenamefont
  {Schneider}, \citenamefont {Moreira}, \citenamefont {Andrade}, \citenamefont
  {Havlin},\ and\ \citenamefont {Herrmann}}]{schneider2011mitigation}%
  \BibitemOpen
  \bibfield  {author} {\bibinfo {author} {\bibfnamefont {C.~M.}\ \bibnamefont
  {Schneider}}, \bibinfo {author} {\bibfnamefont {A.~A.}\ \bibnamefont
  {Moreira}}, \bibinfo {author} {\bibfnamefont {J.~S.}\ \bibnamefont
  {Andrade}}, \bibinfo {author} {\bibfnamefont {S.}~\bibnamefont {Havlin}}, \
  and\ \bibinfo {author} {\bibfnamefont {H.~J.}\ \bibnamefont {Herrmann}},\
  }\href@noop {} {\bibfield  {journal} {\bibinfo  {journal} {Proceedings of the
  National Academy of Sciences}\ }\textbf {\bibinfo {volume} {108}},\ \bibinfo
  {pages} {3838} (\bibinfo {year} {2011})}\BibitemShut {NoStop}%
\bibitem [{\citenamefont {Zhou}\ \emph {et~al.}(2019)\citenamefont {Zhou},
  \citenamefont {L{\"u}},\ and\ \citenamefont {Mariani}}]{zhou2019fast}%
  \BibitemOpen
  \bibfield  {author} {\bibinfo {author} {\bibfnamefont {F.}~\bibnamefont
  {Zhou}}, \bibinfo {author} {\bibfnamefont {L.}~\bibnamefont {L{\"u}}}, \ and\
  \bibinfo {author} {\bibfnamefont {M.~S.}\ \bibnamefont {Mariani}},\
  }\href@noop {} {\bibfield  {journal} {\bibinfo  {journal} {Communications in
  Nonlinear Science and Numerical Simulation}\ }\textbf {\bibinfo {volume}
  {74}},\ \bibinfo {pages} {69} (\bibinfo {year} {2019})}\BibitemShut {NoStop}%
\bibitem [{\citenamefont {Dong}\ \emph {et~al.}(2021)\citenamefont {Dong},
  \citenamefont {Wang}, \citenamefont {Shekhtman}, \citenamefont {Danziger},
  \citenamefont {Fan}, \citenamefont {Du}, \citenamefont {Liu}, \citenamefont
  {Tian}, \citenamefont {Stanley},\ and\ \citenamefont
  {Havlin}}]{dong2021optimal}%
  \BibitemOpen
  \bibfield  {author} {\bibinfo {author} {\bibfnamefont {G.}~\bibnamefont
  {Dong}}, \bibinfo {author} {\bibfnamefont {F.}~\bibnamefont {Wang}}, \bibinfo
  {author} {\bibfnamefont {L.~M.}\ \bibnamefont {Shekhtman}}, \bibinfo {author}
  {\bibfnamefont {M.~M.}\ \bibnamefont {Danziger}}, \bibinfo {author}
  {\bibfnamefont {J.}~\bibnamefont {Fan}}, \bibinfo {author} {\bibfnamefont
  {R.}~\bibnamefont {Du}}, \bibinfo {author} {\bibfnamefont {J.}~\bibnamefont
  {Liu}}, \bibinfo {author} {\bibfnamefont {L.}~\bibnamefont {Tian}}, \bibinfo
  {author} {\bibfnamefont {H.~E.}\ \bibnamefont {Stanley}}, \ and\ \bibinfo
  {author} {\bibfnamefont {S.}~\bibnamefont {Havlin}},\ }\href@noop {}
  {\bibfield  {journal} {\bibinfo  {journal} {Proceedings of the National
  Academy of Sciences}\ }\textbf {\bibinfo {volume} {118}},\ \bibinfo {pages}
  {e1922831118} (\bibinfo {year} {2021})}\BibitemShut {NoStop}%
\bibitem [{\citenamefont {Battiston}\ \emph {et~al.}(2020)\citenamefont
  {Battiston}, \citenamefont {Cencetti}, \citenamefont {Iacopini},
  \citenamefont {Latora}, \citenamefont {Lucas}, \citenamefont {Patania},
  \citenamefont {Young},\ and\ \citenamefont {Petri}}]{battiston2020networks}%
  \BibitemOpen
  \bibfield  {author} {\bibinfo {author} {\bibfnamefont {F.}~\bibnamefont
  {Battiston}}, \bibinfo {author} {\bibfnamefont {G.}~\bibnamefont {Cencetti}},
  \bibinfo {author} {\bibfnamefont {I.}~\bibnamefont {Iacopini}}, \bibinfo
  {author} {\bibfnamefont {V.}~\bibnamefont {Latora}}, \bibinfo {author}
  {\bibfnamefont {M.}~\bibnamefont {Lucas}}, \bibinfo {author} {\bibfnamefont
  {A.}~\bibnamefont {Patania}}, \bibinfo {author} {\bibfnamefont {J.-G.}\
  \bibnamefont {Young}}, \ and\ \bibinfo {author} {\bibfnamefont
  {G.}~\bibnamefont {Petri}},\ }\href@noop {} {\bibfield  {journal} {\bibinfo
  {journal} {Physics Reports}\ }\textbf {\bibinfo {volume} {874}},\ \bibinfo
  {pages} {1} (\bibinfo {year} {2020})}\BibitemShut {NoStop}%
\bibitem [{\citenamefont {Liao}\ \emph {et~al.}(2017)\citenamefont {Liao},
  \citenamefont {Mariani}, \citenamefont {Medo}, \citenamefont {Zhang},\ and\
  \citenamefont {Zhou}}]{liao2017ranking}%
  \BibitemOpen
  \bibfield  {author} {\bibinfo {author} {\bibfnamefont {H.}~\bibnamefont
  {Liao}}, \bibinfo {author} {\bibfnamefont {M.~S.}\ \bibnamefont {Mariani}},
  \bibinfo {author} {\bibfnamefont {M.}~\bibnamefont {Medo}}, \bibinfo {author}
  {\bibfnamefont {Y.-C.}\ \bibnamefont {Zhang}}, \ and\ \bibinfo {author}
  {\bibfnamefont {M.-Y.}\ \bibnamefont {Zhou}},\ }\href@noop {} {\bibfield
  {journal} {\bibinfo  {journal} {Physics Reports}\ }\textbf {\bibinfo {volume}
  {689}},\ \bibinfo {pages} {1} (\bibinfo {year} {2017})}\BibitemShut {NoStop}%
\bibitem [{\citenamefont {Colizza}\ \emph {et~al.}(2006)\citenamefont
  {Colizza}, \citenamefont {Flammini}, \citenamefont {Serrano},\ and\
  \citenamefont {Vespignani}}]{colizza2006detecting}%
  \BibitemOpen
  \bibfield  {author} {\bibinfo {author} {\bibfnamefont {V.}~\bibnamefont
  {Colizza}}, \bibinfo {author} {\bibfnamefont {A.}~\bibnamefont {Flammini}},
  \bibinfo {author} {\bibfnamefont {M.~A.}\ \bibnamefont {Serrano}}, \ and\
  \bibinfo {author} {\bibfnamefont {A.}~\bibnamefont {Vespignani}},\
  }\href@noop {} {\bibfield  {journal} {\bibinfo  {journal} {Nature physics}\
  }\textbf {\bibinfo {volume} {2}},\ \bibinfo {pages} {110} (\bibinfo {year}
  {2006})}\BibitemShut {NoStop}%
\bibitem [{\citenamefont {Freeman}(1977)}]{freeman1977set}%
  \BibitemOpen
  \bibfield  {author} {\bibinfo {author} {\bibfnamefont {L.~C.}\ \bibnamefont
  {Freeman}},\ }\href@noop {} {\bibfield  {journal} {\bibinfo  {journal}
  {Sociometry}\ }\textbf {\bibinfo {volume} {40}},\ \bibinfo {pages} {35}
  (\bibinfo {year} {1977})}\BibitemShut {NoStop}%
\bibitem [{\citenamefont {Avrachenkov}\ \emph {et~al.}(2008)\citenamefont
  {Avrachenkov}, \citenamefont {Litvak},\ and\ \citenamefont
  {Pham}}]{avrachenkov2008singular}%
  \BibitemOpen
  \bibfield  {author} {\bibinfo {author} {\bibfnamefont {K.}~\bibnamefont
  {Avrachenkov}}, \bibinfo {author} {\bibfnamefont {N.}~\bibnamefont {Litvak}},
  \ and\ \bibinfo {author} {\bibfnamefont {K.~S.}\ \bibnamefont {Pham}},\
  }\href@noop {} {\bibfield  {journal} {\bibinfo  {journal} {Internet
  Mathematics}\ }\textbf {\bibinfo {volume} {5}},\ \bibinfo {pages} {47}
  (\bibinfo {year} {2008})}\BibitemShut {NoStop}%
\bibitem [{Note1()}]{Note1}%
  \BibitemOpen
  \bibinfo {note} {Https://github.com/ FFrankyy/FINDER}\BibitemShut {NoStop}%
\bibitem [{\citenamefont {Martin}\ \emph {et~al.}(2014)\citenamefont {Martin},
  \citenamefont {Zhang},\ and\ \citenamefont
  {Newman}}]{martin2014localization}%
  \BibitemOpen
  \bibfield  {author} {\bibinfo {author} {\bibfnamefont {T.}~\bibnamefont
  {Martin}}, \bibinfo {author} {\bibfnamefont {X.}~\bibnamefont {Zhang}}, \
  and\ \bibinfo {author} {\bibfnamefont {M.}~\bibnamefont {Newman}},\
  }\href@noop {} {\bibfield  {journal} {\bibinfo  {journal} {Physical review
  E}\ }\textbf {\bibinfo {volume} {90}},\ \bibinfo {pages} {052808} (\bibinfo
  {year} {2014})}\BibitemShut {NoStop}%
\bibitem [{\citenamefont {Bianconi}(2017)}]{bianconi2017fluctuations}%
  \BibitemOpen
  \bibfield  {author} {\bibinfo {author} {\bibfnamefont {G.}~\bibnamefont
  {Bianconi}},\ }\href@noop {} {\bibfield  {journal} {\bibinfo  {journal}
  {Physical Review E}\ }\textbf {\bibinfo {volume} {96}},\ \bibinfo {pages}
  {012302} (\bibinfo {year} {2017})}\BibitemShut {NoStop}%
\bibitem [{Note2()}]{Note2}%
  \BibitemOpen
  \bibinfo {note} {Http://konect.cc/networks/}\BibitemShut {NoStop}%
\bibitem [{\citenamefont {Dorogovtsev}(2010)}]{dorogovtsev2010complex}%
  \BibitemOpen
  \bibfield  {author} {\bibinfo {author} {\bibfnamefont {S.}~\bibnamefont
  {Dorogovtsev}},\ }\href@noop {} {\emph {\bibinfo {title} {Complex
  networks}}}\ (\bibinfo  {publisher} {Oxford University Press},\ \bibinfo
  {year} {2010})\BibitemShut {NoStop}%
\bibitem [{\citenamefont {Pastor-Satorras}\ and\ \citenamefont
  {Vespignani}(2001)}]{pastor2001epidemic}%
  \BibitemOpen
  \bibfield  {author} {\bibinfo {author} {\bibfnamefont {R.}~\bibnamefont
  {Pastor-Satorras}}\ and\ \bibinfo {author} {\bibfnamefont {A.}~\bibnamefont
  {Vespignani}},\ }\href@noop {} {\bibfield  {journal} {\bibinfo  {journal}
  {Physical review letters}\ }\textbf {\bibinfo {volume} {86}},\ \bibinfo
  {pages} {3200} (\bibinfo {year} {2001})}\BibitemShut {NoStop}%
\bibitem [{\citenamefont {Newman}(2002)}]{newman2002spread}%
  \BibitemOpen
  \bibfield  {author} {\bibinfo {author} {\bibfnamefont {M.~E.}\ \bibnamefont
  {Newman}},\ }\href@noop {} {\bibfield  {journal} {\bibinfo  {journal}
  {Physical review E}\ }\textbf {\bibinfo {volume} {66}},\ \bibinfo {pages}
  {016128} (\bibinfo {year} {2002})}\BibitemShut {NoStop}%
\bibitem [{\citenamefont {Shu}\ \emph {et~al.}(2015)\citenamefont {Shu},
  \citenamefont {Wang}, \citenamefont {Tang},\ and\ \citenamefont
  {Do}}]{shu2015numerical}%
  \BibitemOpen
  \bibfield  {author} {\bibinfo {author} {\bibfnamefont {P.}~\bibnamefont
  {Shu}}, \bibinfo {author} {\bibfnamefont {W.}~\bibnamefont {Wang}}, \bibinfo
  {author} {\bibfnamefont {M.}~\bibnamefont {Tang}}, \ and\ \bibinfo {author}
  {\bibfnamefont {Y.}~\bibnamefont {Do}},\ }\href@noop {} {\bibfield  {journal}
  {\bibinfo  {journal} {Chaos: An Interdisciplinary Journal of Nonlinear
  Science}\ }\textbf {\bibinfo {volume} {25}},\ \bibinfo {pages} {063104}
  (\bibinfo {year} {2015})}\BibitemShut {NoStop}%
\end{thebibliography}
%


\appendix
\section{Baseline methods}
In the section, we introduce the state-of-the-art heuristic and analytical algorithms which are widely used to identify the eigenshield nodes in complex networks. Heuristic methods are based on intuitions about what being a central node means, and therefore, they do not
optimize an objective influence function.
By contrast, analytical methods usually optimize some objective functions to derive the importance of nodes.

\subsubsection{Heuristic methods}

(a) High-Degree (HD): In the HD method, nodes are ranked by degree, and sequentially chosen by the descending order of degree. The drawback of HD is that some hubs may form tightly-knit groups called ``rich-club" \cite{colizza2006detecting} and HD is inclined to choose the rich-club hubs. Whereas rich-club hubs have much overlapping influence, which limits their set influence.

(b) Betweenness centrality (BC):  The betweenness of a node is determined by the number of the shortest paths that pass through the node \cite{freeman1977set}. The betweenness method selects vital nodes according to the descending order of nodes' betweenness. Compared with HD, BC could identify some sparse, yet vital nodes. In this work, we use \textit{MatlabBGL} toolbox 
to implement the BC centrality.

(c) PageRank (PR):
PageRank index was introduced by Brin and Page \cite{brin1998anatomy} to rank web pages in Google's Web search engine, 
and subsequently found diverse applications across biology, scientometrics, and so on~\cite{liao2017ranking}.
For an undirected network, the PageRank scores of nodes are calculated by the recursive equation,
 \begin{equation}\label{eq:pagerank}
\mathbf{s_{t+1}}=\alpha \mathsf{P}\mathbf{s_t}+(1-\alpha)\mathbf{v},
 \end{equation}
 where $\mathsf{P}_{ij}=a_{ij}/d_j$ is the transition matrix. $\mathbf{v}$ is the teleportation vector that is tuned by the parameter $\alpha$. In the experiments, we set $\alpha=0.85$ \cite{liao2017ranking,avrachenkov2008singular} and $\mathbf{v}=\mathbf{1}$.
We  implement the PageRank and set the initial $\mathbf{s_0}=\mathbf{1}$.  We iterate Eq. \ref{eq:pagerank} until the  $|\mathbf{s_{t+1}}-\mathbf{s_t}|$ is less than 0.01\%.


(d) Eigenvector centrality (EC): It utilizes the eigenvector corresponding to the largest
eigenvalue of the adjacency matrix. The importance of a node is characterized by the corresponding entry of the eigenvector.
Nodes are chosen according to the descending order of the entries of the eigenvector.
We directly calculate the eigenvector corresponding to the largest eigenvalue and choose nodes based on the entries of eigenvector.


(e) K-shell: K-shell ranking is based on the K-shell decomposition of the network~\cite{kitsak2010identification}.
The importance of a node is the K-shell of the node. The K-shell of a node is calculated by a recursive procedure: Removing the nodes with degree less than $k'$ iteratively until node $i$ is removed. The minimal $k'$ that node $i$ is removed is the K-shell of the node. K-shell performs well in identifying a single vital node. For multiple nodes, high K-shell nodes also form ``rich-clubs" that have much inner connections between vital nodes, which limits the performance.
In this work, we use our own implementation of the K-shell.

(f) FINDER: This is a deep reinforcement learning method \cite{fan2020finding}. The method is trained purely on small synthetic networks and then applied to real networks. At every step, the method chooses a candidate node that minimizes an accumulated normalized connectivity that depends on the applications. In the experiments, we use the trained reinforcement neural model that is released by the authors \cite{fan2020finding} (see the Github responsibility \footnote{https://github.com/
FFrankyy/FINDER}). The trained reinforcement neural model provides an open API to calculate the vital nodes.



\subsubsection{Analytical methods}
(g) Non-backtracking matrix (NBM): The method utilizes the non-backtracking matrix of a network to evaluate the importance of a node. The non-backtracking matrix was once used to detect communities in sparse networks \cite{krzakala2013spectral}. Comparing with the EC method that causes most of the weight of the centrality to concentrate on a small number of nodes, NBM avoids the concentration of the weight on a small fraction of nodes. the non-backtracking matrix of a network is $\mathsf{P}_{2m\times 2m}$, with the element ${\mathsf{P}}_{i\leftarrow j,k\leftarrow l}=\delta_{jk}(1-\delta_{il})$, where $m$ is the number of directed edges. $\delta_{jk}=1$ if $j=k$; $\delta_{jk}=0$ otherwise. 
The centralities of nodes by NBM are equal to the first $n$ elements of the
leading eigenvector of the $2n\times 2n$ matrix \cite{krzakala2013spectral,martin2014localization},
\begin{equation}
\label{eq:nonback1}
\mathbf{M}=\begin{pmatrix}
              \mathsf{A} & I-D \\
              I & 0 \\
            \end{pmatrix},
\end{equation}
 where $\mathsf{A}$, $I$, $D$ are the adjacency matrix, identity matrix, and the diagonal degree matrix of a network, respectively. Equation \ref{eq:nonback1} provides a convenient access to evaluate the importance of nodes based on non-backtracking matrix.  We implement the Eq. \ref{eq:nonback1} to calculate the importance of nodes.


(h) Collective Influence (CI) \cite{morone2015influence}:
Morone and Makse proposed the collective influence that utilized the non-backtracking matrix to find the optimal vital nodes. After approximately solving the percolation problem, they arrive at a simple formalism to choose vital nodes.
The method  assigns to node $i$ the collective influence
strength,
\begin{equation}
\label{eq:CIdefinition}
CI_\ell(i)=(k_i-1)\sum_{j\in \partial Ball(i,\ell)}(k_j-1),
\end{equation}
where $\partial Ball(i,\ell)$ is the set of nodes with $\ell-$length from node $i$. Vital nodes are chosen based on the CI of each node. The method introduces a tunable parameter $\ell$ to tune the sphere of influence of a node.
We implement the CI to calculate the importance of each node. The $CI_\ell(i)$ of all remaining nodes are re-calculated when we chose a node and add it into the eigenshield node set.


(i) Belief propagation method (BP): NBM and CI assume that information does not spread back to the previous nodes, and hence neglect the loop structures. A more accurate method is to utilize the belief propagation method \cite{mugisha2016identifying} or (message-passing method \cite{bianconi2017fluctuations}). The BP algorithm is rooted in the spin glass model. In BP, a recursive process is adopted to calculate the importance of nodes. The probability $q_i^0(t)$ that a node $i$ is suitable to be removed from the network is
\begin{equation}
\label{eq:bp}
q_i^0=\frac{1}{1+e^x[1+\sum_{k\partial i(t)}\frac{1-q_{k\rightarrow i}^0}{q_{k\rightarrow i}^0+q_{k\rightarrow i}^k}]\prod_{j\in \partial i(t)}[q_{j\rightarrow i}^0+q_{j\rightarrow i}^j]},
\end{equation}
where $x$ is a tunable parameter and $\partial i(t)$ is the set of neighboring nodes of node $i$ at time $t$. $q_{j\rightarrow i}^0(t)$ is the probability that the neighboring node $j$ is suitable to be removed if node $i$ is absent at time $t$, and while $q_{j\rightarrow i}^j(t)$ is the probability that the neighboring node $j$ is suitable to be the root node of a tree-like component if node $i$ is absent at time $t$. The two marginal probability values $q_{j\rightarrow i}^0(t)$ and $q_{j\rightarrow i}^j(t)$ are estimated by a self-consistent belief propagation equation:
\begin{equation}
\label{eq:bp1}
q_{j\rightarrow i}^0=\frac{1}{z_{j\rightarrow i}(t)},
\end{equation}
\begin{equation}
\label{eq:bp2}
q_{j\rightarrow i}^i=\frac{e^x\prod_{k\in \partial i(t)\setminus j}[q_{k\rightarrow i}^0+q_{k\rightarrow i}^j]}{z_{j\rightarrow i}(t)},
\end{equation}
where $\partial i(t)\setminus j$ is the set of neighboring nodes of node $i$ excluding $j$. $z_{j\rightarrow i}(t)$ is a normalization operation as
\begin{eqnarray}
\label{eq:bp3}
z_{j\rightarrow i}(t)=1+e^x&&\prod_{k\in \partial i(t)\setminus j}[q_{k\rightarrow i}^0+q_{k\rightarrow i}^j]\times \nonumber\\
&&[1+\sum_{l\in \partial i(t)\setminus j}\frac{1-q_{l\rightarrow i}^0}{q_{l\rightarrow i}^0+q_{l\rightarrow i}^l}].
\end{eqnarray}

In order to obtain a fixed number of vital nodes, at every step, we iterate Eqs. \ref{eq:bp1}--\ref{eq:bp3} to the stable state and then obtain $q_i^0$ by Eq. \ref{eq:bp}. The node with the highest $q_i^0$ is removed from the network. We repeat the iteration process again and obtain only one vital node at each step.

We implement the descrete-time BP: Initially, we set $q_{j\rightarrow i}^i(t=0)=0.5$. At every step, we use set $q_{j\rightarrow i}^i(n)$ to update $q_{j\rightarrow i}^i(n+1)$ and $q_{j\rightarrow i}^0(n+1)$. The iteration ends when the $|q_{j\rightarrow i}^i(n+1)-q_{j\rightarrow i}^i(n)|<0.01\%$ for all edge $(i,j)$.

\section{Data description}
The empirical datasets are all from the  konect dataset collection \footnote{http://konect.cc/networks/}. We treat all networks unweighted and undirected. Besides, we only reserve the giant component of a network to ensure that the network is connected. The details of all networks are as follows:

(1) Facebook ego network: This network consists of `circles' (or `friends lists') from Facebook, which was collected from survey participants using Facebook app. The data has been anonymized by replacing the Facebook-internal ids for each user with a new value. 

(2) Reality Mining: This undirected network contains human contact data among 100 students of the Massachusetts Institute of Technology (MIT), collected by the Reality Mining experiment performed in 2004 as part of the Reality Commons project. A node represents a person and an edge indicates that the corresponding nodes had physical contact.

(3) Dbpedia-similar: This is the similarity graph from DBpedia. It contains the ``similar to" links between pages of Wikipedia. The network is undirected and does not contain multiple edges.

(4) Gene fusion: This is a gene fusion network. Nodes are genes, and two nodes are connected if the two genes have been observed to have fused during the emergence of cancers.

(5) PDZBase: This is a network of protein-protein interactions from PDZBase.

(6) Jazz: The collaboration network between Jazz musicians. Each node is a Jazz musician and an edge denotes whether two musicians have played together in a band in 2003.

(7) Haggle: The undirected network represents contacts between people measured by carried wireless devices. A node represents a person and an edge shows the contact between persons.

(8) Netscience: This is a network of co-authorships in the area of network science.

(9) Infectious: This network describes the face-to-face behavior of people during the exhibition INFECTIOUS: STAY AWAY in 2009 at the Science Gallery in Dublin. Nodes represent exhibition visitors and edges represent face-to-face contacts that were active for at least 20 seconds.

(10) Elegans: This is the metabolic network of the roundworm Caenorhabditis elegans. Nodes are metabolites (e.g., proteins), and edges are interactions between them.

(11) aS7332: The graph represents the Internet Autonomous Systems (AS) topology. Each AS exchanges traffic flows with some neighbors (peers). The network was constructed from the BGP (Border Gateway Protocol) logs of the University of Oregon Route Views Project - Online data and reports from November 8, 1997 to January 2, 2000.

(12) Euroroads: This is the international E-road network located mostly in Europe. The network is undirected. Nodes represent cities and an edge between two nodes denotes that they are connected by an E-road.

(13) Arenas-email: This is the email communication network at the University Rovira i Virgili in Tarragona in the south of Catalonia in Spain. Nodes are users and each edge represents that at least one email was sent.

(14) Air traffic: This network was opened by the USA's FAA (Federal Aviation Administration) National Flight Data Center (NFDC), Preferred Routes Database. Nodes in this network represent airports or service centers and links are created from strings of preferred routes recommended by the NFDC.

(15) Yeast: This undirected network contains protein interactions in yeast. A node represents a protein and an edge represents a metabolic interaction between two proteins.

(16) Hamsterster friendships: This Network contains friendships between users of the website hamsterster.com.

(17) DNC emails: This is the directed network of emails in the 2016 Democratic National Committee (DNC) email leak. Nodes in the network correspond to persons in the dataset. A directed edge in the dataset denotes that a person has sent an email to another person.

(18) Human protein (Stelzl): This network represents interacting pairs of protein in Humans.

(19) US power grid: This undirected network contains information about the power grid of the Western States of the United States of America. An edge represents a power supply line. A node is either a generator, a transformator or a substation.

(20) Bitcoin:
This is who-trusts-whom network of people who trade using Bitcoin on a platform called Bitcoin OTC. Members of Bitcoin OTC rate other members in a scale of -10 (total distrust) to +10 (total trust) in steps of 1. This is the first explicit weighted signed directed network available for research.

(21) Route views: This is the undirected network of autonomous systems of the Internet connected with each other. Nodes are autonomous systems (AS), and edges denote communication.

(22) WikiVote: Wikipedia is a free encyclopedia written collaboratively by volunteers around the world. A small part of Wikipedia contributors are administrators. In order for a user to become an administrator, a request for adminship (RfA) is issued and the Wikipedia community proposes a public discussion or a vote deciding who to promote to adminship.
The network contains all the Wikipedia voting data from the inception of Wikipedia till January 2008. Nodes in the network represent wikipedia users and a directed edge from node $i$ to node $j$ represents that user $i$ voted on user $j$.

(23) CaHepTh:
Arxiv HEP-TH (High Energy Physics - Theory) collaboration network is from the e-print arXiv and covers scientific collaborations between authors papers submitted to High Energy Physics - Theory category. If an author $i$ co-authored a paper with author $j$, the graph contains an undirected edge from $i$ to $j$. If the paper is co-authored by $k$ authors this generates a completely connected (sub)graph on $k$ nodes.
The data covers papers in the period from January 1993 to April 2003 (124 months).

(24) Sister cities: This is an undirected network of cities of the world connected by ``sister city" or ``twin city" relationships extracted from WikiData.

(25) Oregon: This is the Autonomous Systems (AS) peering information inferred from Oregon route-views between March 31, 2001 and May 26, 2001.

(26) Astrophysics: This is the co-authorship network from the "astrophysics" section (astro-ph) of arXiv. Nodes are authors and an edge denotes a collaboration.

(27) Douban: This is the social network of douban, a Chinese online recommendation site. The network is undirected and unweighted.

(28) GoogleHyperlink:
This is a network of web pages connected by hyperlinks. The data was released in 2002 by Google as a part of the Google Programming Contest.

(29) CAIDA: This is the undirected network of autonomous systems of the Internet from the CAIDA project, collected in 2007. Nodes are autonomous systems (AS), and edges denote communication.

(30) Digg: This is the reply network of the social news website Digg. Each node in the network is a user of the website, and each directed edge denotes that a user replied to another user.

(31) Amazon (MDS):
This is the co-purchase network of Amazon based on the ``customers who bought this also bought" feature. Nodes are products and an undirected edge between two nodes shows that the corresponding products have been frequently bought together.

(32) Brightkite: This undirected network contains user-user friendship relations from Brightkite website. A node represents a user.

(33) Catster households: This is an undirected online social network.

(34) Livemocha: This is the social network of Livemocha, an online language learning community. The network is undirected and unweighted.

(35) CiteSeer: This is the citation network extracted from the CiteSeer digital library. Nodes are publications and the directed edges denote citations.

(36) Actor collaborations: This is an actor network. Two actors are connected  if they both appeared in the same movie.

(37) Dogster: This Network contains friendships between users of the website dogster.com.

(38) Youtube friendship:
This is the friendship network of the video-sharing site Youtube. Nodes are users and an undirected edge between two nodes indicates a friendship.

(39) Hyves: This is the social network of Hyves, a Dutch online social network. The network is undirected and unweighted.

(40) Orkut: This is the social network of Orkut users and their connections. The dataset is only a sample of the orkut website and  thus may be incomplete.

(41) Erd\"{o}s-R\'{e}nyi (ER) network:  This is a random network generated by the ER model \cite{dorogovtsev2010complex}.

(42-46) Scale free (SF) network:  This is a random network following the power law degree distribution. Initially, we randomly determine the degree of each node based on the degree distribution $p(d)=Ck^{-\gamma}$ and then randomly connect the nodes \cite{dorogovtsev2010complex}.

\begin{table*}
\caption{Structural properties of the different real networks. Structural properties include network size ($N$), link number ($E$), degree assortativity ($r$), average clustering coefficient ($<C>$), average shortest path length
($<L>$) and sparsity. }
\label{tab:dataset}
\begin{center}
\begin{longtable}{p{4cm} p{1.6cm}  p{2.1cm} p{1.2cm} p{1.5cm} p{1.5cm}p{2cm}}
\hline
\hline
 Network&$N$ &$E$ &$r$ &$<C>$ &$<L>$& Sparsity\\
\hline
1 Facebook           &94 &187 &-0.673 &0.006 &3.412&$4.28\times10^{-2}$\\
2 Reality Mining &96&2593&-0.056& 0.597&1.445&$5.56\times10^{-1}$\\
3 Dbpedia-similar &107&167&-0.140&0.037& 5.212&$2.94\times10^{-2}$\\
4 Gene fusion &110&124&-0.454&-0.000& 4.248&$2.07\times10^{-2}$\\
5 PDZBase          &161&209&-0.466&0.001&5.326&$1.62\times10^{-2}$\\
6 Jazz          &197&2719&0.020&0.449&2.234&$1.41\times10^{-1}$\\
7 Haggle        &273&2048&-0.481&0.089&2.424&$5.51\times10^{-2}$\\
8 Netscience  &379&914&-0.082&0.146&6.042&$1.28\times10^{-2}$\\
9 Infectious     &409&2760&0.231&0.327&3.632&$3.31\times10^{-2}$\\
10 Elegans        &452&2021&-0.225&0.096&2.664&$1.98\times10^{-2}$\\
11 aS7332          &492 &1078  &-0.236 &0.027 &3.434&$8.92 \times10^{-3}$\\
12 Euroroads  &1039&1305&0.090&0.005&18.395&$2.40\times10^{-3}$\\
13 Arenas-email &1133&5451&0.078&0.084&3.606&$8.50 \times10^{-3}$\\
14  Air traffic    &1225&2399&-0.016&0.0115&5.932&$3.20\times10^{-3}$\\
15 Yeast  &1458&1948&-0.210&0.010&6.812&$1.80\times10^{-3}$\\
16 Hamsterster &1788&12476&-0.089&0.028&3.453&$7.80\times10^{-3}$\\
17 DNC emails     &1864&4267&-0.304&0.079&3.365&$2.46\times10^{-3}$\\
18 Human protein   &3022&6104&-0.123&0.012&4.861&$1.34\times10^{-3}$\\
19 US power grid    &4940&6591&0.004&0.015&18.990&$5.40\times10^{-4}$\\
20 Bitcoin        &5880&21228&-0.163&0.022&3.587&$1.23\times10^{-3}$\\
21 Route views &6474&13895&-0.182&0.010&3.667&$6.63\times10^{-4}$\\
22 WikiVote        &7066&100736&-0.083&0.040&3.248&$4.03\times10^{-3}$\\
23 CaHepTh        &8638&24806&0.239&0.141&5.945&$6.65\times10^{-4}$\\
24 Sister cities &14274&20573&0.387 &0.111&	7.654&$2.02\times10^{-4}$\\
25 Oregon        &10670&22002&-0.186&0.006&3.642&$3.87\times10^{-4}$\\
26 Astrophysics &16046&121251&0.235&0.425&5.108&$9.42\times10^{-4}$\\
27 Douban        &154907 &327103  &-0.180 &0.010 &5.103&$2.73\times10^{-5}$\\
28 GoogleHyperlink  &15762&137184&-0.122&0.192&2.561&$1.10\times10^{-3}$\\
29 CAIDA &26475&53381&-0.238&0.007&3.912&$2.44\times10^{-4}$\\
30 Digg        &30359 &85146  &0.005 &0.006 &4.682&$1.85\times10^{-4}$\\
31 Amazon        &334862 &925864  &-0.059 &0.205 &11.731&$1.65\times10^{-5}$\\
32 Brightkite &58228 &214078&0.011&	0.111&4.859&$1.26\times10^{-4}$\\
33 Catster households 	&105138 &494858&-0.134&0.004&2.617&$8.95\times10^{-5}$\\
34 Livemocha  &104103 &	2193083&-0.147&0.014&3.207&$4.05\times10^{-4}$\\
35 CiteSeer &384413 &1751463&-0.061&0.050&	6.348&$2.37\times10^{-5}$\\
36 Actor collaborations 	&382219 &33115812&0.227&0.166&3.698&$4.53\times10^{-4}$\\
37 Dogster  &426820 &8546581&-0.088&0.014&3.399&$9.38\times10^{-5}$\\
38 Youtube        &1134889 &2987595  &-0.037 &0.006 &5.554&$4.64\times10^{-6}$\\
39 Hyves  &1402673 &2777419&-0.023&0.002&5.756&$2.82\times10^{-6}$\\
40 Orkut 	 &3072441 &117184899&0.016&0.041&4.267&$2.48\times10^{-5}$\\
41 ER        &5000&40000&-0.008&0.002&2.000&$3.20\times10^{-3}$\\
42 SF, r=4.0        &5000&40000&-0.004&0.005&2.000&$3.20\times10^{-3}$\\
43 SF, r=3.5        &5000&40000&-0.010&0.007&2.000&$3.20\times10^{-3}$\\
44 SF, r=3.0        &5000&40000&-0.0000&0.015&2.000&$3.20\times10^{-3}$\\
45 SF, r=2.5        &5000&40000&-0.0004&0.019&2.000&$3.20\times10^{-3}$\\
46 SF, r=2.0        &5000&40000&0.003&0.023&2.000&$3.20\times10^{-3}$\\
\hline
\hline
\end{longtable}
\vspace*{0.0cm}
\end{center}
\end{table*}

\section{Spreading models and Parameter settings}
\textbf{SIS model}: Considering the susceptible-infected-susceptible (SIS) spreading model \cite{pastor2001epidemic,newman2002spread} in a network denoted by an adjacency matrix $\mathsf{A}=(a_{ij})_{N\times N}$. Let $\rho_i(t)$ represent the infection probability of node $i$ at time $t$.  The general dynamics of each node could be written as \cite{shu2015numerical,pastor2001epidemic}
\begin{equation}\label{eq:hmf}
  \frac{\partial \rho_i(t)}{\partial t}=-\rho_i(t)+\beta  [1-\rho_i(t)]\sum_{j\in N_i}\rho_j(t),
\end{equation}
where $N_i$ is the neighboring set of node $i$.

In the experiment, we consider the SIS immunization process. Initially, a small fraction of eigenshield nodes are chosen and immunized. The chosen nodes are removed from the network and don't participate in the spreading of epidemics. At the same time, 5\% random nodes are set as infected ones and the corresponding $\rho_i(t=0)=1$. We then use discrete-time approach to simulate the SIS model: Time is divided into small uniform steps of a certain duration. $\rho_i(n+1)=\rho_i(n)+[-\rho_i(n)+\beta  (1-\rho_i(n))\sum_{j\in N_i}\rho_j(n)]\Delta t$, where $\Delta t=0.00001$.  The iteration ends when  $|\sum_i\rho_i(n+1)-\sum_i\rho_i(n)|<0.01\%$.


\textbf{SIR model}: In the spread of epidemics, when infected individuals recover, they have the  immunizing power for the epidemics. In order to characterize the phenomenon, recovered state is introduced into the spreading process, which is called SIR model. The dynamics of SIR model is
\begin{eqnarray}
\label{eq:SIRmodel}
\left\{
\begin{aligned}
\frac{dS_v(t)}{dt}&=-\beta S_v(t)\sum_{z\in N_v}I_z(t),\\
\frac{dI_v(t)}{dt}&=\beta S_v(t)\sum_{z\in N_v}I_z(t)-\gamma I_v(t),\\
\frac{dR_v(t)}{dt}&=\gamma I_v(t),
\end{aligned}
\right.
\label{eq:SIR}
\end{eqnarray}
where $S_v(t)$, $I_v(t)$ and $R_v(t)$ mean the susceptible probability, infected probability and recovered probability of node $v$ at time $t$. $\beta$, $\gamma$ mean the infecting rate and recovering rate.

The spreading ability is evaluated by the fraction of nodes that were once infected. That is to say,
\begin{equation}\label{eq:SIRinfectaionfraction}
\tau=\frac{1}{N}\sum_v I_v(t\rightarrow +\infty)+R_v(t\rightarrow +\infty).
\end{equation}


In the experiment, we simulate the SIR immunization process. Initially, a small fraction of eigenshield nodes are chosen and immunized. The chosen nodes are removed from the network and don't participate in the spreading of epidemics. At the same time, 5\% random nodes are set as infected ones and the corresponding $I_v(t=0)=1$ and the other nodes are in susceptible state. We then use discrete-time approach to simulate the SIR model, which is similar to that of SIS model. The iteration ends when  $|\sum_vI_v(n+1)-\sum_vI_v(n)|<0.01\%$ and  $|\sum_vR_v(n+1)-\sum_vR_v(n)|<0.01\%$.

\textbf{Linear threshold model}:
In the model, a node $i$ is infected by its neighbor $j$ based on the weight of edge $b_{ij}$, which requires $\sum_{j\in N_i}b_{ij}\leq1$. For undirected networks, $b_{ij}$
 is usually defined  as $b_{ij}=1/d_i$, where $d_i$ is the degree of node $i$. The dynamics of linear threshold model \cite{kempe2003maximizing} is as follows:

1. Initially, all nodes are in inactive state.

2. Assign a threshold $\theta_i$ for each node $i$.

3. Choose a small number of nodes as initial spreaders and set them as active state. 

4. In step $t$, all nodes that were active in step $t-1$ remain active. In the meanwhile, we activate the nodes whose total weight of their active neighbors is larger than the threshold, i.e.,
\begin{equation}\label{eq:linearthreshold}
\sum_{N_{i,activated}}b_{ij}\geq \theta_i,
\end{equation}
The threshold $\theta_i$ indicates the intrinsic tendencies of nodes to adopt the action of its active neighbors. 

5. Repeat step 4 until no fresh node is activated in the step.

In the maximization problem of linear threshold model, the key issue is to choose effective initial active nodes to maximize the size of activated nodes in the ultimate step, denoted by $\sigma(\mathsf{S})$, where $\mathsf{S}$ is the initial spreaders.

\textbf{Parameter settings}:
For the spreading immunization simulations of SIS and SIR models, we set the spreading rate $\beta = 0.3$ and 5\% random nodes as initial infected nodes unless otherwise stated.
For the SIR model, the recovering rate is settled $ \gamma= 0.1$ and 5\% random nodes is settled as initial infected ones. Actually, we also investigate the spreading rate $\beta$ on the SIR model.
For the linear threshold model, $\theta_i$ usually follow random distribution. In the experiments, we use uniform distribution $(0,1)$ for $\theta_i$ 
in Eq. \ref{eq:linearthreshold} in the experiments.
 In order to reduce the fluctuation, we run 100 independent simulation to obtain the average size of the activated nodes in the ultimate step.

When choosing eigenshield nodes, some baseline methods has tunable parameters. For the PR method, we set $\alpha=0.85$ and $\mathbf{v}=1$ in Eq. \ref{eq:pagerank} because the settings achieve best performance for the PageRank method. For the CI method, we set $\ell=2$, because if $\ell$ is large, $CI_\ell(i)=0$ in Eq. \ref{eq:CIdefinition}  in small networks, and under the scenario we cannot evaluate the importance of most nodes. $\ell=2$ could achieve quite good performance in both small and large networks. For the BP method, we set the initial parameters in Eqs. \ref{eq:bp}--\ref{eq:bp} following uniform distribution $(0,1)$.
For the proposed OI method, we set $h=20$ in Eq. \ref{eq:setinfluence} unless otherwise stated (In the experiments, we also investigate the influence of $h$ on the performance).

\section{Additional experiments}
\begin{figure*}[htbp]
  \centering
  \includegraphics[width=6in,angle=0,trim=0 0 0  0,clip=true]{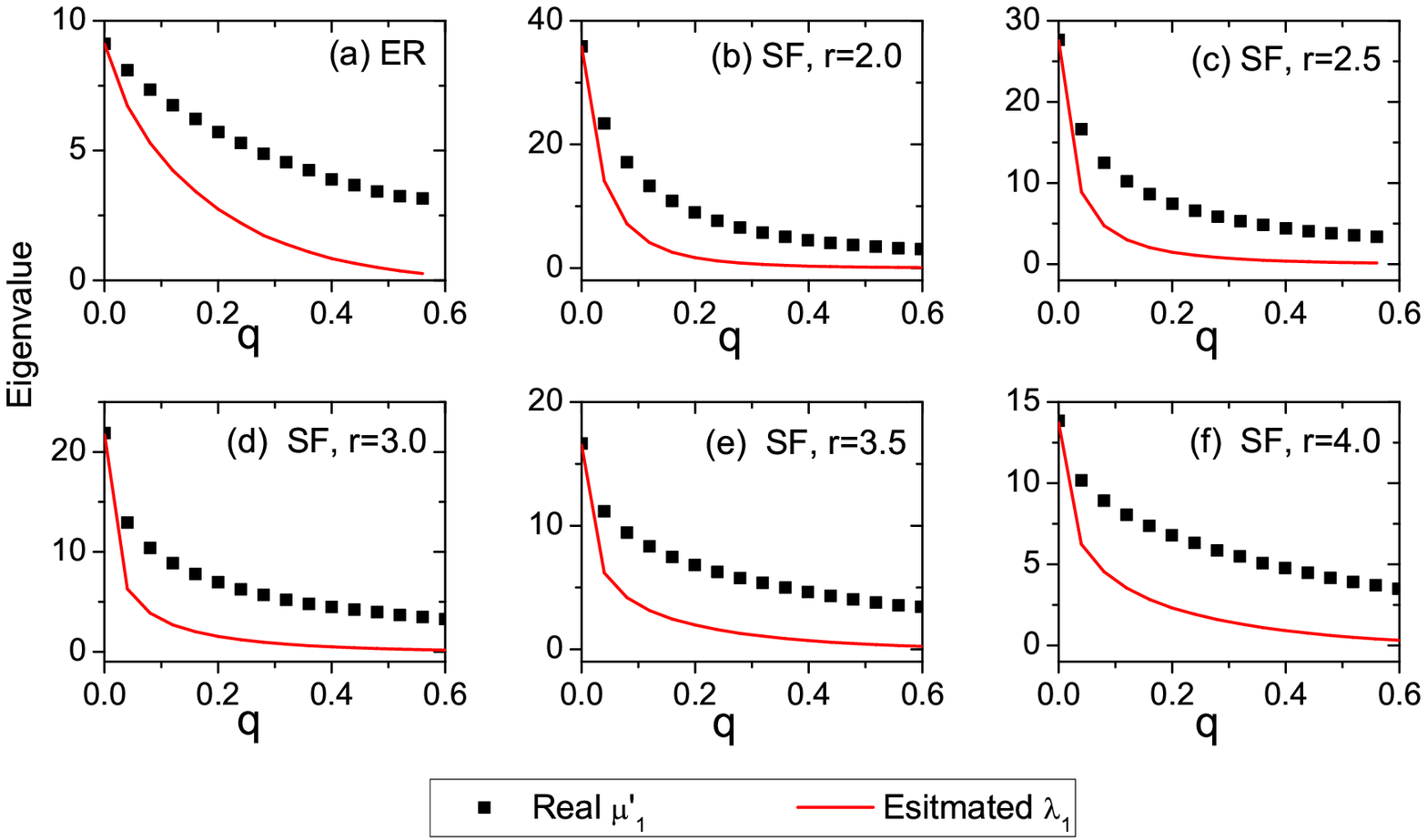}
  \caption{(Color online) The difference between real and estimated largest eigenvalues of the remaining matrix $\mathsf{A}'$ in model networks as a function of the fraction $q$ of the eigenshield nodes. The eigenshield nodes are chosen by high degree (HD) method.
   (a) ER network. (b) SF network, $\gamma=2.0$. (c) SF network, $\gamma=2.5$. (c) SF network, $\gamma=3.0$. (d) SF network, $\gamma=3.5$. (e) SF network, $\gamma=4.0$.
  }
  \label{fig:Approximatemodellambda}
\end{figure*}

\begin{figure*}[htbp]
  \centering
  \includegraphics[width=6in,height=8in,angle=0,trim=0 0.6in 0  0,clip=true]{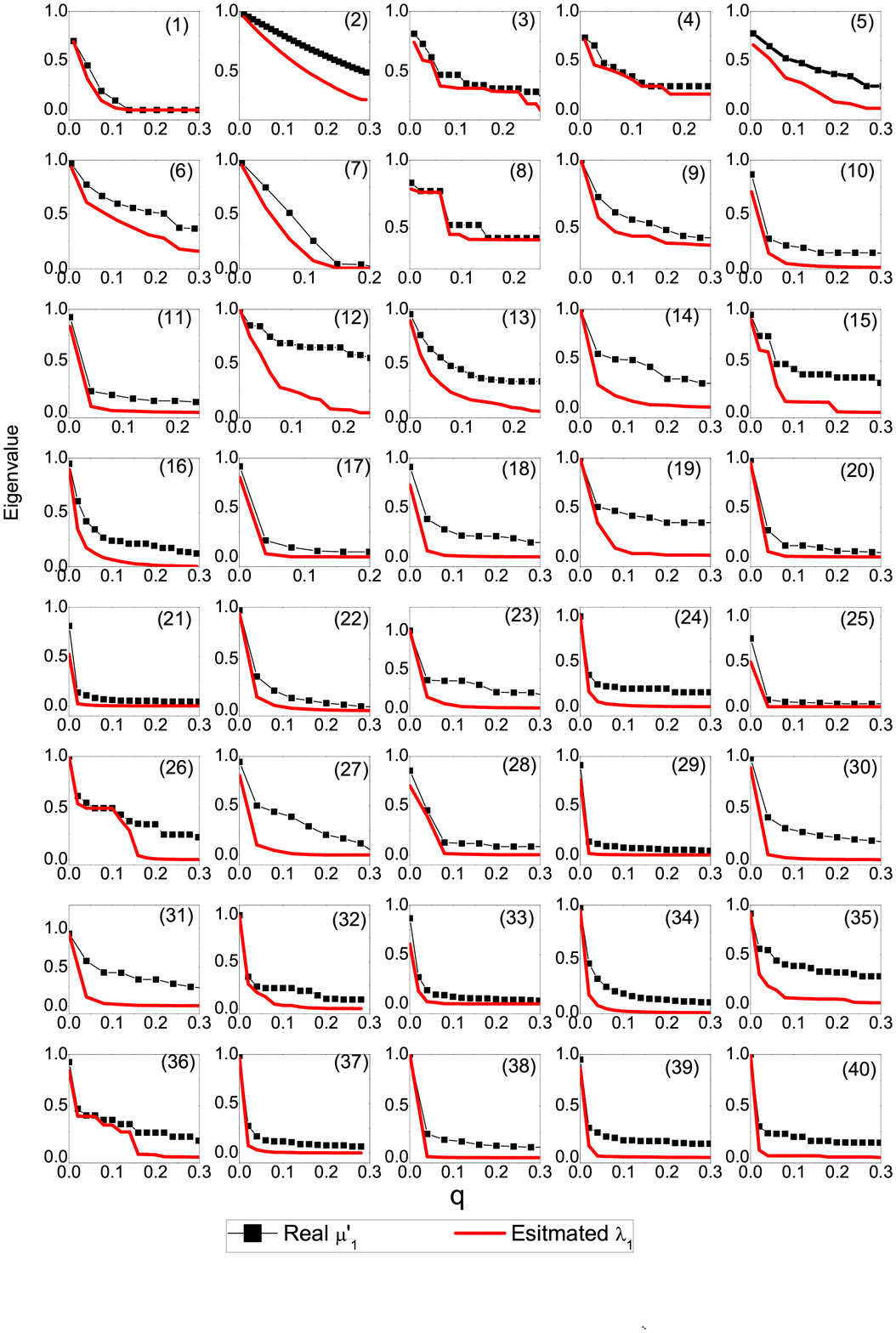}
  \caption{(Color online) The difference between real and estimated largest eigenvalues of the remaining matrix $\mathsf{A}'$ in real networks. Panels $(1)\sim(40)$ are the results of the 40 real networks, following the order of table I.
  }
  \label{fig:Approximatereallambda}
\end{figure*}

Figure \ref{fig:Approximatemodellambda} shows the difference between real and estimated largest eigenvalues, where the estimated largest eigenvalue is obtained by Eq. \ref{eq:setinfluence}. In Fig. \ref{fig:Approximatemodellambda}, the difference is small when the fraction $q$ of the eigenshield nodes is very small and then keeps stable for large $q$. The real networks also have the similar performance in Fig. \ref{fig:Approximatereallambda}. Hence, the validity of Eq. \ref{eq:basic} is empirically illustrated for small $q$.

\begin{figure*}[htbp]
  \centering
  \includegraphics[width=4in,trim=0 0in 0in  0,clip=true]{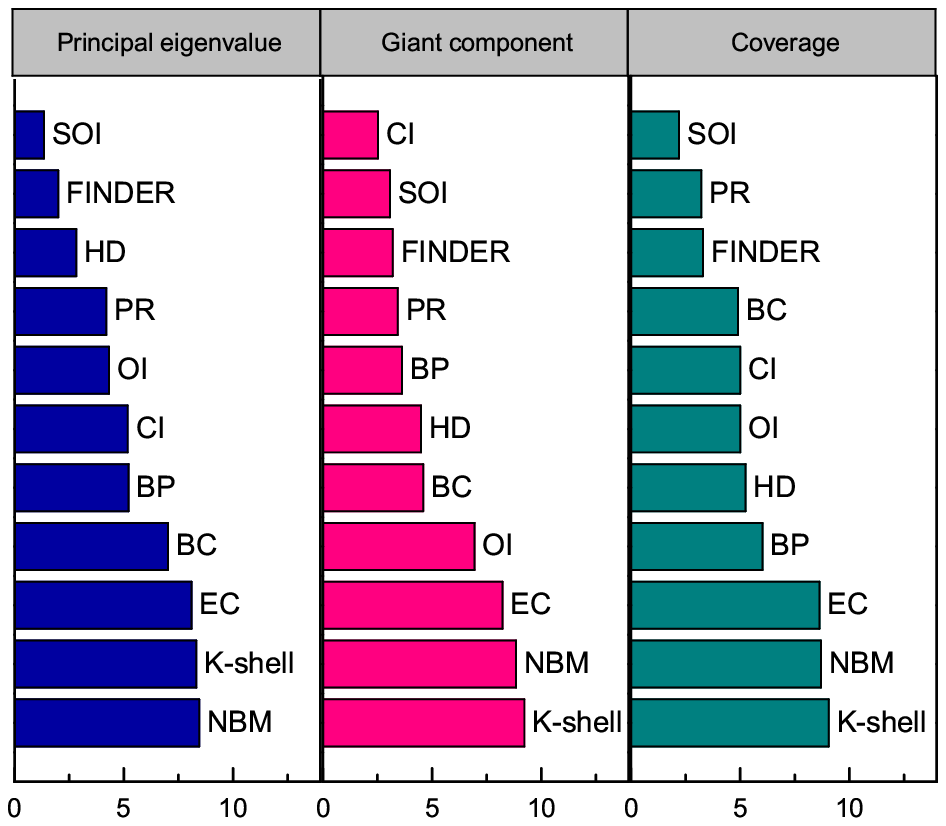}\\
  \caption{(Color online) The means of ranking orders of the eleven methods based on the three metrics $R_{\mu_1}$, $R_{G}$ and $R_{\sigma}$.
We use ascending order for the $R_{\mu_1}$ and $R_{G}$, and descending order for $R_{\sigma(q)}$. Thus, smaller order is better.
  }
  \label{fig:ranking}
\end{figure*}

Figure \ref{fig:ranking} shows the average ranking order of the eleven methods based on the three metrics $R_{\mu_1}$, $R_{G}$ and $R_{\sigma}$. We see that the SOI performs the best in terms of $R_{\mu_1}$ and $R_{\sigma}$, which agrees well with the performance of the average ranking score.

\begin{figure*}[htbp]
  \centering
  \includegraphics[width=4in,angle=0,trim=0 0 0  0,clip=true]{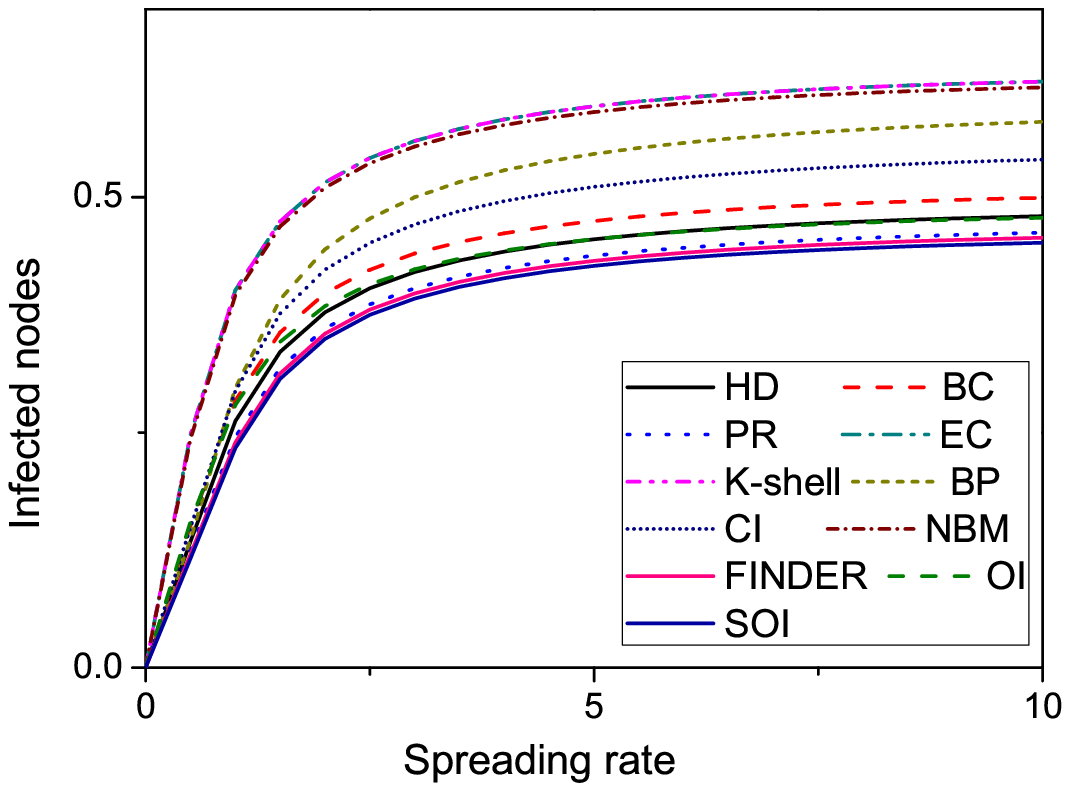}
  \caption{(Color online) The average fraction of infected nodes as a function of the spreading rate for the SIS spreading model across 40 real networks.
   We simulate the SIS immunization process: Initially, a small fraction of eigenshield nodes are chosen and immunized. The chosen nodes are removed from the network and don't participate in the spreading of epidemics. At the same time, 5\% random nodes are set as initial infected ones and the other nodes are in susceptible state. We count the fraction of infection nodes when  $|\sum_i\rho_i(n+1)-\sum_i\rho_i(n)|<0.01\%$. Smaller fraction of infected nodes means better immunization performance and is better.}

  \label{fig:LTSummarization}
\end{figure*}

Note that SIS model and linear threshold model are nonlinear dynamics and the dynamics is determined by both the location of eigenshield nodes and the spreading rate. 
In the experiments, we choose 10\% nodes as eigenshield nodes. For the SIS model, we remove the eigenshield nodes from the network and that removing the eigenshield nodes could hinder the spreading of information. For the linear threshold model, we set the eigenshield nodes as initially activated ones.
  We then observe the influence of the spreading rate on the final activated nodes for the two models. The average performances of each method across the 40 real networks are summarized in Figs. \ref{fig:LTSummarization} and \ref{fig:rankingscore}(c). 
  Based on the performance summary, the SOI method outperforms other methods irrespective of the spreading rate.

\begin{figure*}[htbp]
  \centering
  \includegraphics[width=4in,angle=0,trim=0 0 0  0,clip=true]{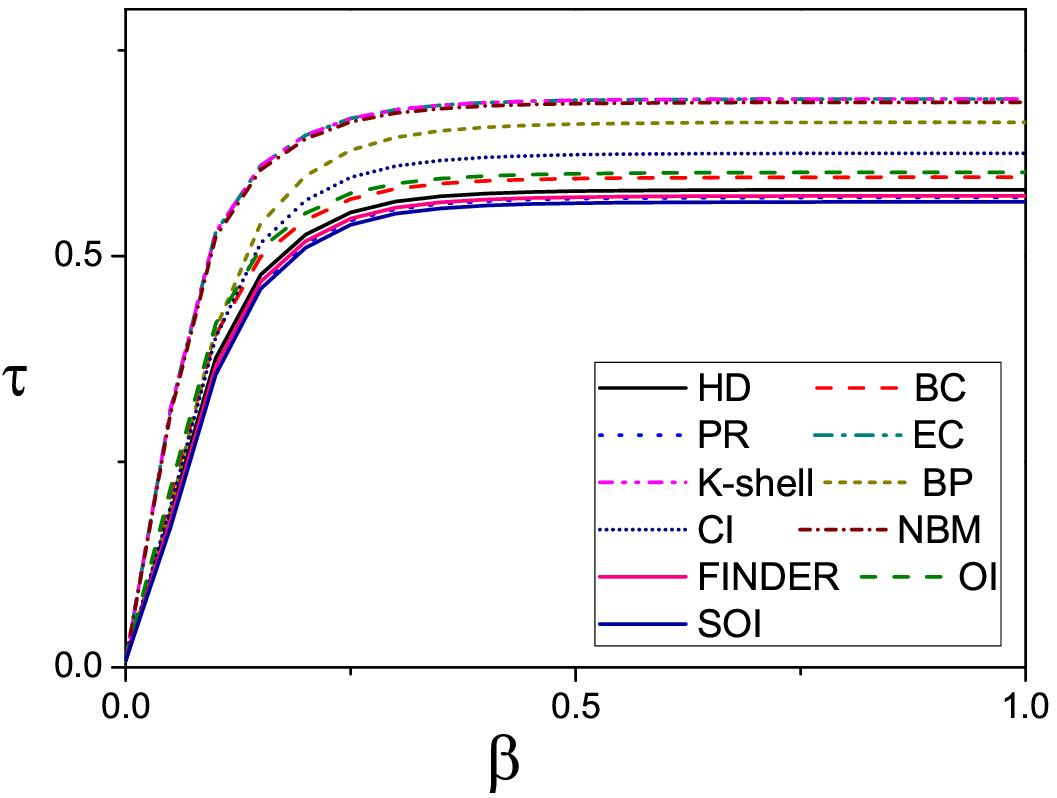}
  \caption{(Color online) The average fraction of infection nodes and recovering nodes as a function of the spreading rate $\beta$ (in Eq. \ref{eq:SIRmodel}) for the SIR spreading model across 40 real networks.
   We simulate the SIR immunization process: Initially, a small fraction of eigenshield nodes are chosen and immunized. The chosen nodes are removed from the network and don't participate in the spreading of epidemics. At the same time, 5\% random nodes are set as initial infected ones and the other nodes are in susceptible state, $\gamma=0.1$. We count the fraction of infection nodes and recovering nodes when  $|\sum_vI_v(n+1)-\sum_vI_v(n)|<0.01\%$ and  $|\sum_vR_v(n+1)-\sum_vR_v(n)|<0.01\%$. Smaller $\tau$ means better immunization performance and is better.}
  \label{fig:SIRsummarizationinfectionrate}
\end{figure*}

\begin{figure*}[htbp]
  \centering
  \includegraphics[width=4in, angle=0,trim=0 0 0  0,clip=true]{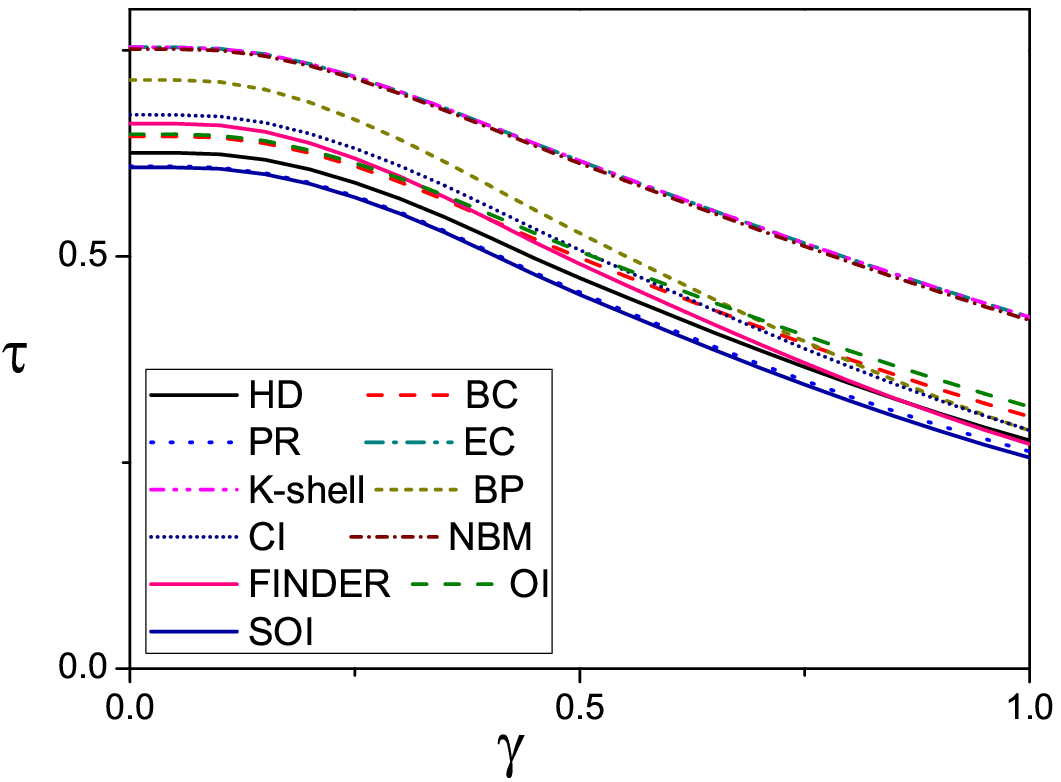}
  \caption{(Color online) The average fraction of infection nodes and recovering nodes as a function of the recovering rate $\beta$ (in Eq. \ref{eq:SIRmodel}). In the simulation, we set the spreading rate $\beta=0.5$ in Eq. \ref{eq:SIRmodel}. The simulation process is similar to that of Fig. \ref{fig:SIRsummarizationinfectionrate}.
  }
  \label{fig:SIRsummarizationrecoverate}
\end{figure*}

Besides, we also study the influence blocking problem based on the SIR epidemic spreading model.  10\% nodes are chosen as eigenshield ones and are removed from the network. The removal of the eigenshield nodes could hinder the spreading of information. The average performances of each method across the 40 real networks are summarized in Figs. \ref{fig:SIRsummarizationinfectionrate} and \ref{fig:SIRsummarizationrecoverate}. Similar to the SIS model, the SOI method also outperforms other methods irrespective of the parameter settings.


\begin{figure*}[htbp]
  \centering
  \includegraphics[width=4in,height=3.5in,angle=0,trim=0 0 0  0,clip=true]{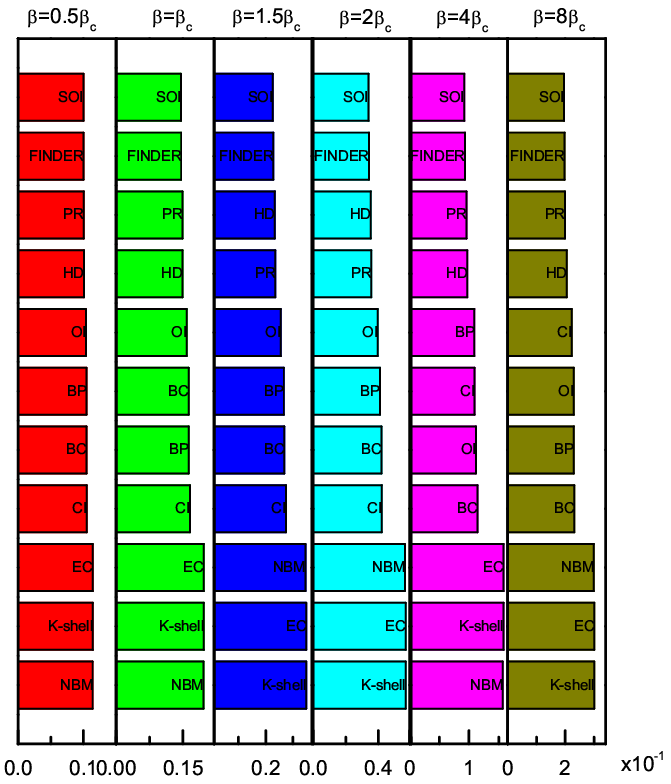}
  \caption{(Color online) The average fraction of infected nodes as a function of the spreading rate for the SIS model across 40 real networks. Smaller is better.  
   The spreading rates are $\beta=0.5\beta_c,\beta_c,1.5\beta_c,2\beta_c,4\beta_c,8\beta_c$ respectively, where $\beta_c=1/\mu_1$.
  }
  \label{fig:net40SISsimulation}
\end{figure*}

\begin{figure*}[htbp]
  \centering
  \includegraphics[width=4in,height=3.5in, angle=0,trim=0 0 0  0,clip=true]{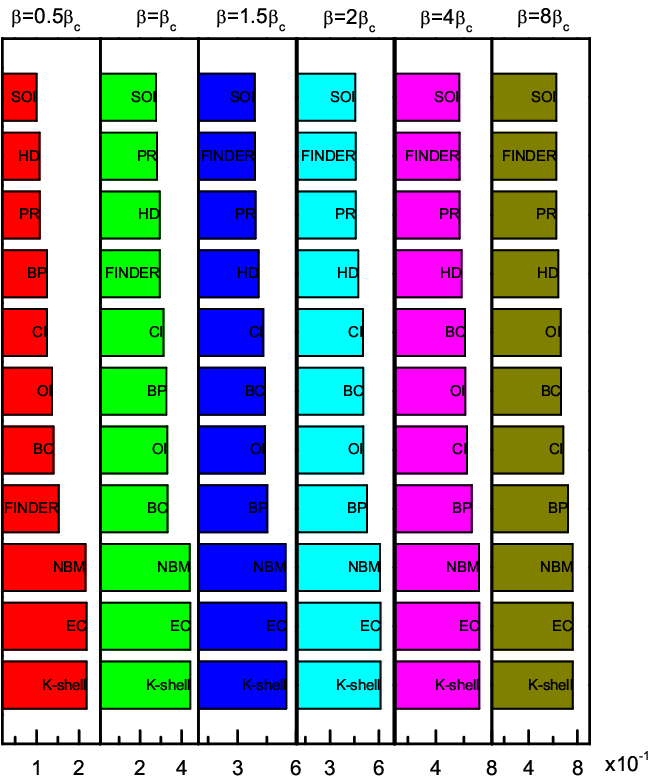}
  \caption{(Color online) The average fraction of infected nodes as a function of the spreading rate (in Eq. \ref{eq:SIRmodel}) for the SIR spreading model across 40 real networks. The recovering rate is set as $\gamma=0.1$ in Eq. \ref{eq:SIRmodel}. Smaller fraction of infected nodes is better.  
  The spreading rate is $\beta=0.5\beta_c,\beta_c,1.5\beta_c,2\beta_c,4\beta_c,8\beta_c$ respectively, where $\beta_c=1/\mu_1$.
  }
  \label{fig:net40SIRsimulation}
\end{figure*}

The percolation of SIS and SIR in complex networks reveals that the spreading dynamics dramatically change near the threshold (We use $\beta_c=1/\mu_1$ to characterize the threshold \cite{chakrabarti2008epidemic}). We particularly simulate the dynamics with different spreading rate below, near, and above the threshold. Figures \ref{fig:net40SISsimulation} and \ref{fig:net40SIRsimulation} shows that the SOI method always performs the best irrespective of the spreading rate.


\begin{figure*}
  \includegraphics[width=5in]{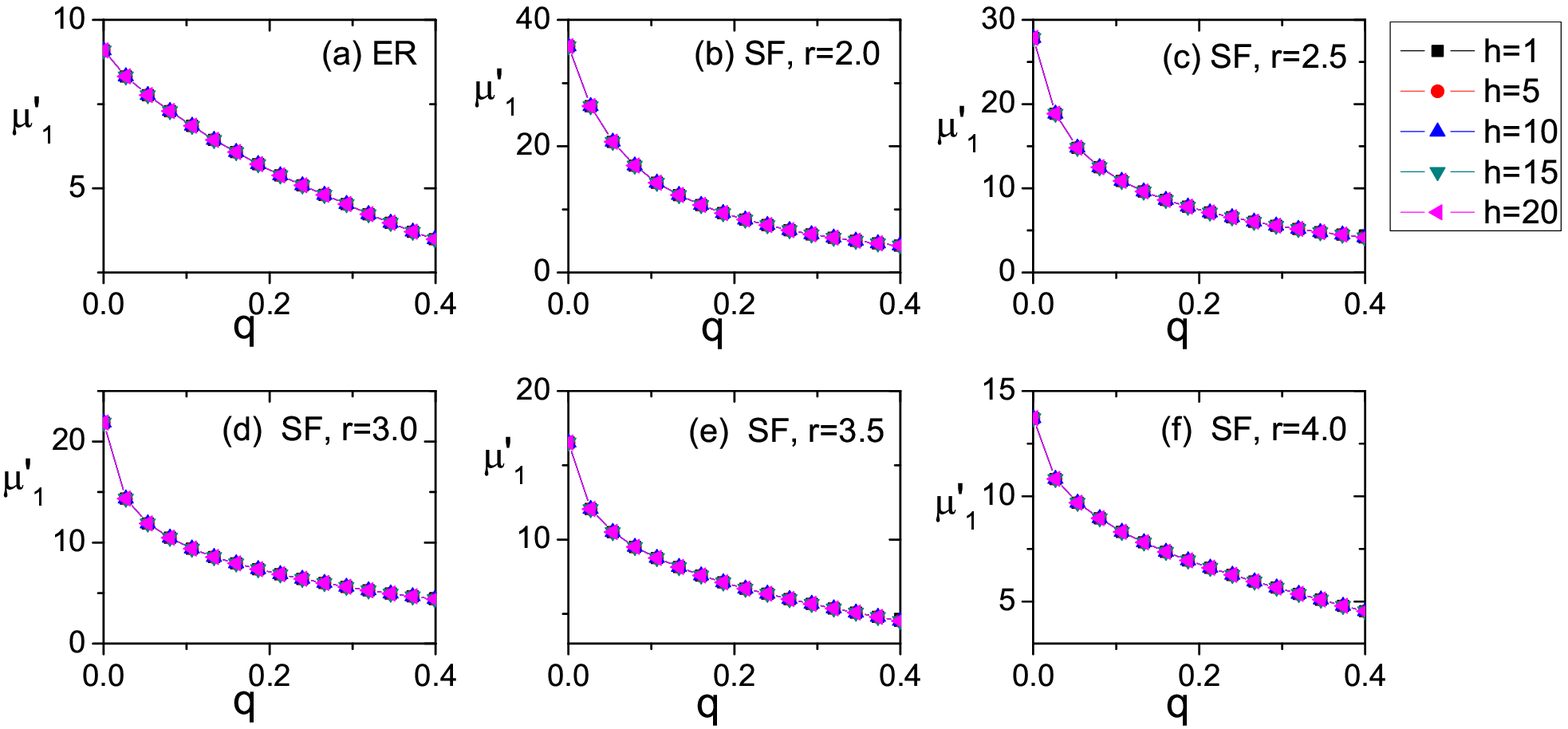}
  \caption{(Color online) The influence of $h$ on the eigenvalue performance of the proposed method OI in model networks.  (a) ER network. (b) SF network, $\gamma=2.0$. (c) SF network, $\gamma=2.5$. (c) SF network, $\gamma=3.0$. (d) SF network, $\gamma=3.5$. (e) SF network, $\gamma=4.0$.
  }
  \label{fig:dimensionmodellambda}
\end{figure*}

\begin{figure*}
  \centering
  \includegraphics[width=6.5in,height=8in,angle=0,trim=0 0.6in 0  0,clip=true]{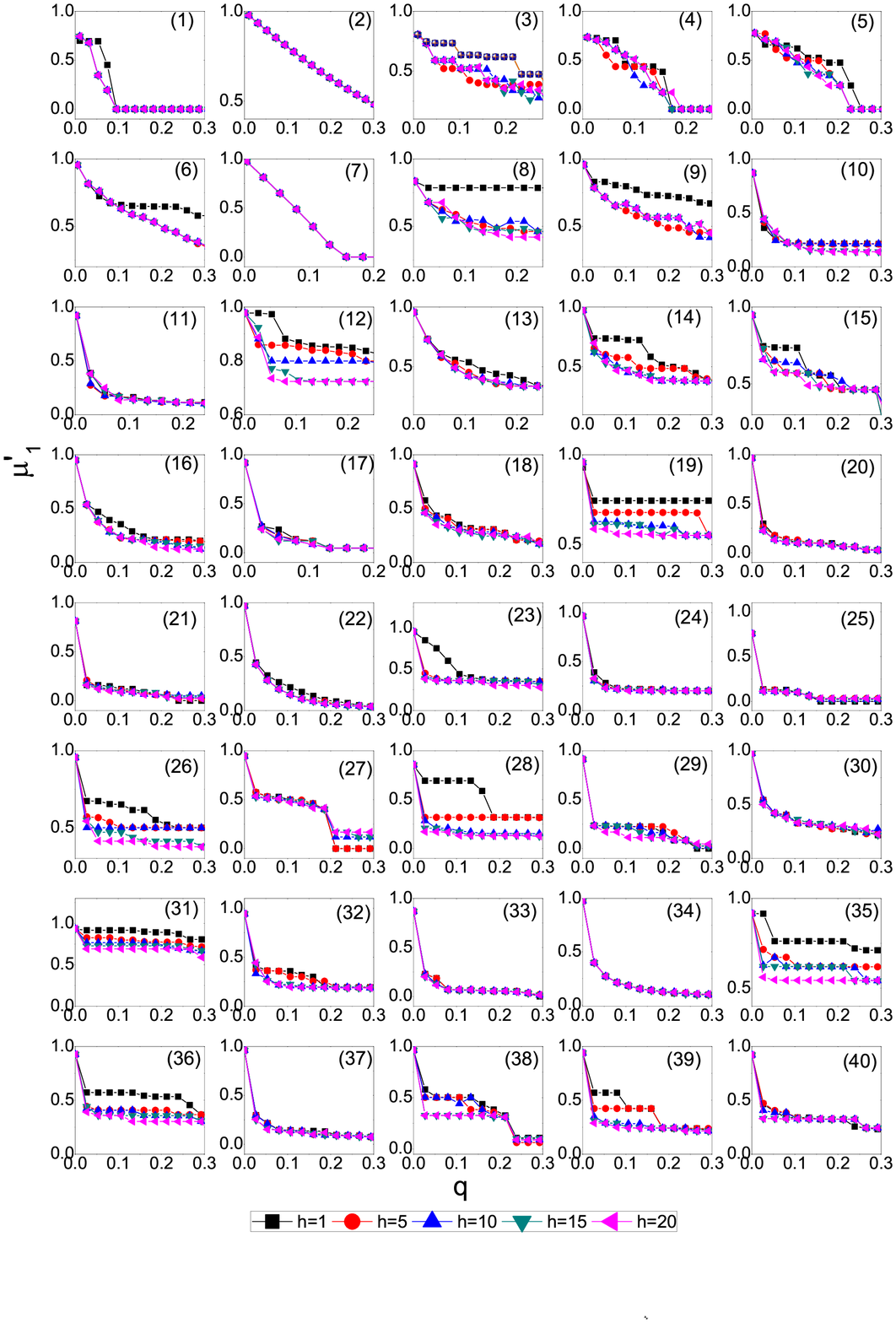}
  \caption{(Color online) The influence of $h$ on the eigenvalue performance of the proposed method OI in real networks.  Panels $(1)\sim(40)$ are the results of the 40 real networks, following the order of table I.
  }
  \label{fig:dimensionreallambda}
\end{figure*}

Further, Figures \ref{fig:dimensionmodellambda} and \ref{fig:dimensionreallambda} show the influence of $h$ on the performance of the proposed method OI in model networks and real networks respectively.
In mode networks (Fig. \ref{fig:dimensionmodellambda}), $h$ rarely influences the eigenvalue, while the performance increases with $h$ in real networks, since real networks have complex network structures that are difficult to characterize. However large $h$ would increase the time consumption, and thus we use $h=20$ in the experiments.

\begin{figure*}
  \centering
  \includegraphics[width=6in]{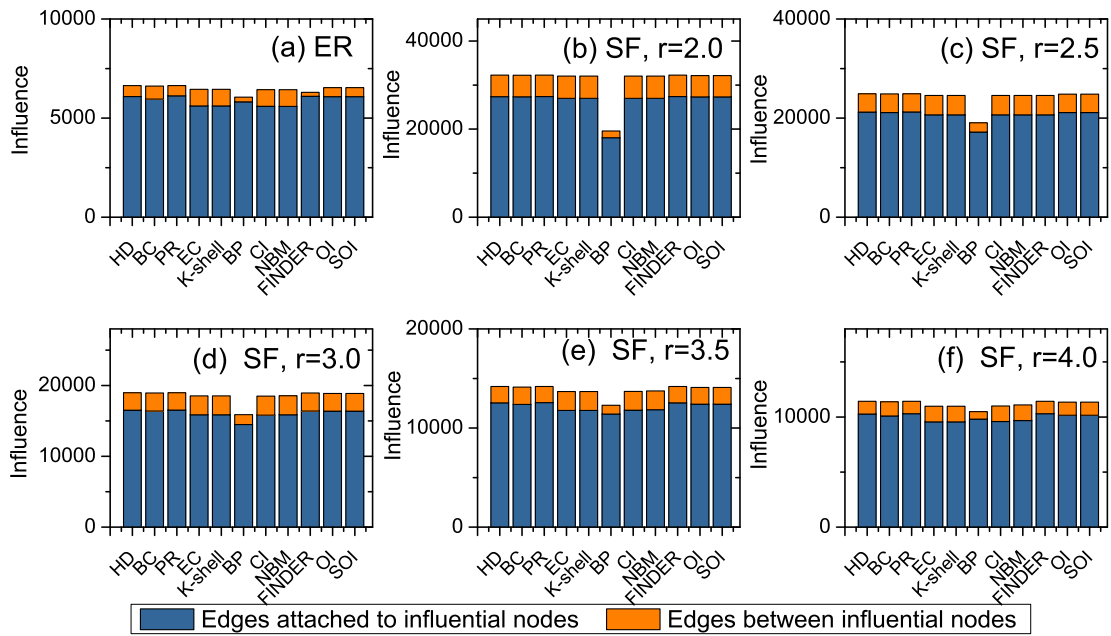}
  \caption{(Color online)  The size of edges attached to eigenshield nodes and edges between eigenshield nodes in model networks when $q=0.1$.  (a) ER network. (b) SF network, $\gamma=2.0$. (c) SF network, $\gamma=2.5$. (c) SF network, $\gamma=3.0$. (d) SF network, $\gamma=3.5$. (e) SF network, $\gamma=4.0$.
  }
  \label{fig:overlapmodeldegree}
\end{figure*}

\begin{figure*}
  \centering
  \includegraphics[width=6.5in,height=8in,angle=0,trim=0 0in 0  0,clip=true]{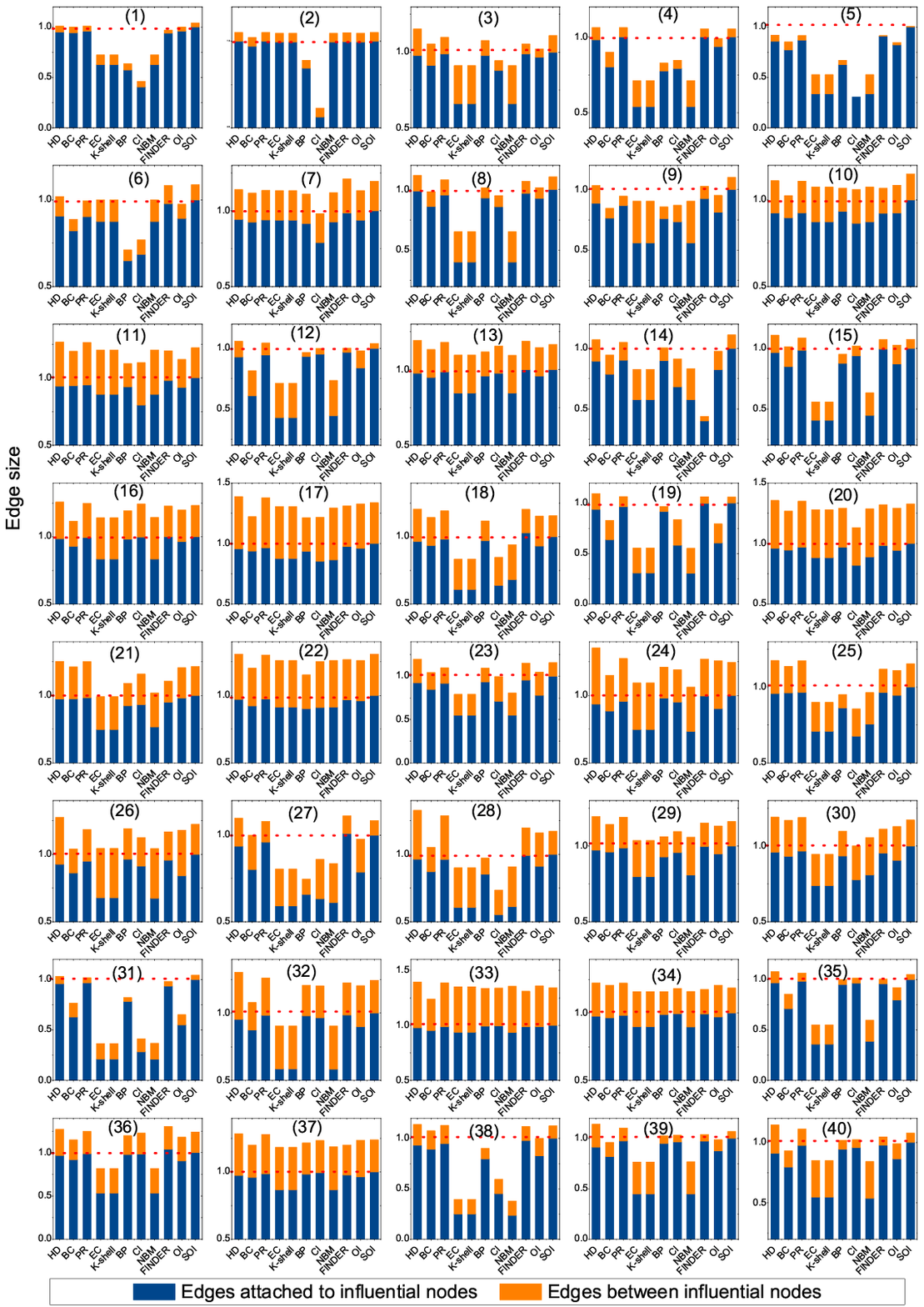}
  \caption{(Color online)  The size of edges attached to eigenshield nodes and edges between eigenshield nodes in real networks when $q=0.1$.  Panels $(1)\sim(40)$ are the results of the 40 real networks, following the order of table I.
  }
  \label{fig:overlaprealdegree}
\end{figure*}

Moreover, we analyze the size of edges attached to eigenshield nodes and edges between eigenshield nodes in model and real networks when $q=0.1$ in Figs. \ref{fig:overlapmodeldegree} and \ref{fig:overlaprealdegree}.  In model networks, since most methods are inclined to choose high degree nodes, the performance of different methods is similar on the whole. Whereas in real networks, we see that the SOI method has the most edges attached to eigenshield nodes, but the least edges between eigenshield nodes, and thus, the set influence is maximized and overlapping influence is minimized.
\clearpage


\section{Matrix perturbation theory}
Equation \ref{eq:basic} is the basis of the proposed method.
In the section, we show the derivation detail of Eq. \ref{eq:basic}.

 We now consider the matrix perturbation on arbitrary matrix.
Supposing that the matrix $\mathsf{A}$ has left and right eigenvectors, denoted as $ \mathbf{y}_1, \mathbf{y}_2,..., \mathbf{y}_n$ and $\mathbf{v}_1,\mathbf{v}_2,...,\mathbf{v}_n$, and that all the left and right eigenvalues are simple, i.e., without multiplicity, we first define quantities $s_i$,
 \begin{equation}
s_i=\mathbf{y}_i^T\mathbf{v}_i \quad (i=1,2,...,n).
\end{equation}
Note that $s_i$ actually represents the cosine similarity of $\mathbf{y}_i$ and $\mathbf{v}_i$ if the two vectors are real and normalized. When $||\mathbf{y}_i||_2=1$ and $||\mathbf{v}_i||_2=1$, we also have
 \begin{equation}
|s_i|=|\mathbf{y}_i^T\mathbf{v}_i|\leq ||\mathbf{y}_i||_2||\mathbf{v}_i||_2=1.
\end{equation}

We now consider the eigenvalues $\mu_i(\epsilon)$ and eigenvectors $\mathbf{v}_i(\epsilon)$ of matrix $\mathsf{A}+\epsilon \mathsf{P}$.
We denote $t_{ij}$ as parameters. Since $\mu_i(\epsilon)$ and $\mathbf{v}_i(\epsilon)$ can be represented as power series, substituting
 \begin{equation}
 \mu_i(\epsilon)=\mu_i+k_{i,1}\epsilon+k_{i,2}\epsilon^2+...,
 \end{equation}
 and
 \begin{equation}
\mathbf{v}_{i}(\epsilon)=\mathbf{v}_{i}+(\epsilon t_{11}+\epsilon^2 t_{12}+...)\mathbf{v}_{1}+...+(\epsilon t_{n1}+\epsilon^2 t_{n2}+...)\mathbf{v}_{n}
 \end{equation}
into the
 \begin{equation}
 \label{eq:eqeig}
(\mathsf{A}+\epsilon \mathsf{P})\mathbf{v}_{i}(\epsilon)=\mu_i(\epsilon)\mathbf{v}_{i}(\epsilon),
\end{equation}
we get
 \begin{equation}
 \label{eq:firstterms1}
\mathsf{A}(\sum_{j=1,j\neq i}^n t_{ji}\mathbf{v}_j)+\mathsf{P}\mathbf{v}_i=\mu_i(\sum_{j=1,j\neq i}^n t_{ji}\mathbf{v}_j)+k_{i,1}\mathbf{v}_i.
\end{equation}
The Eq. \ref{eq:firstterms1} can be simplified as
 \begin{equation}
 \label{eq:firstterms2}
\sum_{j=1,j\neq i}^n (\mu_j-\mu_i)t_{ji}\mathbf{v}_j+\mathsf{P}\mathbf{v}_i=k_{i,1}\mathbf{v}_i.
\end{equation}
Left-multiplying the equation by $\mathbf{y}_i^T$ and notice that $\mathbf{y}_i^T\mathbf{v}_j=0$,($i\neq j$), we get
 \begin{equation}
 \label{eq:eigvalueperturb}
 k_{i,1}=\frac{\mathbf{y}_i^T\mathsf{P}\mathbf{v}_i}{\mathbf{y}_i^T\mathbf{v}_i}.
 \end{equation}


Next, we analyze the perturbation of eigenvector based on Eq. \ref{eq:firstterms2}. Left multiplying Eq. \ref{eq:firstterms2} by $\mathbf{y}_i^T$, we get
 \begin{equation}
(\mu_j-\mu_i)t_{ji}s_j+\beta_{ji}=0\quad (j=1,2,3,...,n, j\neq i).
 \end{equation}
The first term of $\mathbf{v}_i(\epsilon)$ is
 \begin{eqnarray}
\epsilon &&[
\frac{\beta_{1,i}\mathbf{v}_1}{(\mu_i-\mu_1)s_1}+
\frac{\beta_{2,i}\mathbf{v}_2}{(\mu_i-\mu_2)s_2}+...+\frac{\beta_{i-1,i}\mathbf{v}_{i-1}}{(\mu_i-\mu_{i-1})s_{i-1}}+\nonumber\\
&&\frac{\beta_{i+1,i}\mathbf{v}_{i+1}}{(\mu_i-\mu_{i+1})s_{i+1}}+...+
\frac{\beta_{n,i}\mathbf{v}_n}{(\mu_i-\mu_n)s_n}].
 \end{eqnarray}
Note that all other eigenvectors will influence the $\mathbf{v}_i(\epsilon)$. Besides, the influence is also determined by $\mu_i-\mu_j$. If an eigenvalue $\mu_j$ is close to $\mu_i$, the eigenvector corresponding to $\mu_i$ is sensitive to the perturbation.

Given a network, since the adjacency matrix $A$ is symmetrical, we have $\mathbf{y}_i=\mathbf{v}_i$ and $s_i=1$. Hence, we obtain Eq. \ref{eq:basic} based on Eq. \ref{eq:eigvalueperturb}.

\end{document}